\documentclass[11pt]{article}
\usepackage{amsmath}
\usepackage{amsfonts}
\usepackage{latexsym}

\newtheorem{theorem}{Theorem}
\newtheorem{lemma}{Lemma}

\newtheorem{proposition}{Proposition}
\newtheorem{definition}{Definition}
\newtheorem{remark}{{\it Remark}}
\newlength{\dinwidth}
\newlength{\dinmargin}
\def\be{\begin{equation}}
\def\ee{\end{equation}}
\def\beqs{\begin{displaymath}}
\def\eeqs{\end{displaymath}}
\def\beqn{\begin{eqnarray}}
\def\eeqn{\end{eqnarray}}

\def\Lie{{\rm Lie}}

\def\l{\lambda}
\def\I{{\rm I}}
\def\C{\mathbb {C}}
\def\Z{\mathbb {Z}}
\def\R{\mathbb {R}}
\def\N{\mathbb{N}}
\def\F{{\cal F}}

\def\a{\alpha}
\def\b{\beta}
\def\g{\gamma}
\def\d{\partial}
\def\hsp{\hspace{0.5cm}}
\def\L{{\cal L}}

\def\imb{\mathrm{Im}\mathbb B}
\def\B{\mathbb B}
\def\res{\mathrm{res}}
\def\Qbar{\bar{Q}}
\def\Pbar{\bar{P}}
\def\lb{\bar{\l}}

\def\ib{{\bar{i}}}
\def\kb{{\bar{k}}}
\def\nb{{\bar{n}}}
\def\zb{{\bar{z}}}
\def\vp{\mathrm{v.p.}}
\def\jb{{\bar{j}}}
\def\iA{{\scriptscriptstyle{A}}}
\def\iB{{\scriptscriptstyle{B}}}
\def\iC{{\scriptscriptstyle{C}}}
\def\iD{{\scriptscriptstyle{D}}}
\def\iP{{\scriptscriptstyle{P}}}
\def\iI{{\scriptscriptstyle{{\rm I}}}}
\def\iW{{\scriptscriptstyle{W}}}

\def\iL{{\scriptscriptstyle{L}}}
\def\iE{{\scriptscriptstyle{E}}}
\def\iF{{\scriptscriptstyle{F}}}
\def\iPhi{{\scriptscriptstyle{\Phi}}}
\def\iOmega{{\scriptscriptstyle{\Omega}}}
\def\h{{\scriptscriptstyle{(1,0)}}}
\def\ah{{\scriptscriptstyle{(0,1)}}}
\def\i0{{\scriptscriptstyle{0}}}
\def\covM{\widehat{M}}
\def\set{\{1,\dots,L;\bar{1},\dots,\bar{L}\}}
\def \surf {\L}
\def \cov {{\L_\l}}

\def\fpoly{\widehat{\surf}}

\def\z{\varsigma}
\def\p{p}
\def\dim{n}
\def \const {C}

\makeatletter
\@addtoreset{equation}{section}
\makeatother

\begin{document}
\begin{center}
{\Large ``Real doubles" of Hurwitz Frobenius manifolds}
\end{center}
\begin{center}
\vspace{1 cm}

{\large Vasilisa Shramchenko}

\vspace{1 cm}
Department of Mathematics and Statistics, Concordia University\\
7141 Sherbrooke West, Montreal H4B 1R6, Quebec, Canada\\
e-mail: vasilisa@mathstat.concordia.ca
\vspace{1 cm}
\end{center}

\textbf{Abstract.} New Frobenius structures on Hurwitz spaces are found. A Hurwitz space is considered as a real manifold; therefore the number of coordinates is twice as large as the number of coordinates on Hurwitz Frobenius manifolds of Dubrovin. Simple branch points of a ramified covering and their complex conjugates play the role of canonical coordinates on the constructed Frobenius manifolds. Corresponding solutions to WDVV equations and $G$-functions are obtained.


\section{Introduction}
Frobenius manifolds were introduced by B. Dubrovin \cite{Dubrovin0} as a geometric interpretation of the Wit\nolinebreak ten - Dijkgraaf - E.Verlinde - H.Verlinde (WDVV) equations from two-dimensional to\-po\-lo\-gi\-cal field theory \cite{DVV, Witten}. 

The theory of Frobenius manifolds is related to various branches of mathematics: the theory of singularities -- some ingredients of a Frobenius manifold had long existed on the base space of the universal unfolding of a hypersurface singularity. Besides singularity theory, Frobenius manifold structures have been found on cohomology spaces of smooth projective varieties (the theory of Gromov-Witten invariants); on extended moduli spaces of Calabi-Yau manifolds; on  orbit spaces of Coxeter groups, extended affine Weil groups and Jacobi groups; and on Hurwitz spaces (see the references in \cite{Dubrovin, Manin}). 

The aim of the present work is to construct a new class of semisimple (vector algebra on any tangent space has no nilpotents) Frobenius manifolds associated with Hurwitz spaces. The dimension of Dubrovin's Frobenius manifolds on Hurwitz spaces is equal to the complex dimension of the Hurwitz space. In this paper we build Frobenius structures of a double dimension on the real Hurwitz space. We consider the Hurwitz space as a real manifold, i.e. we complement the set of its usual local coordinates by the set of their complex conjugates.  We call new Frobenius manifolds the ``real doubles" of Hurwitz Frobenius manifolds of Dubrovin
(in some cases the prepotential of a ``real double" is  real-valued, however this is not always the case).

We start with a construction of a family of Darboux-Egoroff (flat potential diagonal) metrics on a real Hurwitz space in genus greater than zero. The Hurwitz space we consider is the space of coverings $(\surf,\l)$ of $\C P^1\;,$ where $\surf$ is a Riemann surface of genus $g \geq 1\;,$ $\l$ is a meromorphic function on $\surf$ with simple finite critical points $P_1, \dots, P_\iL$ and possibly with critical points at infinity. The real Hurwitz space has local coordinates $\{ \l_1,\dots,\l_\iL;\lb_1,\dots,\lb_\iL \}\;,$ where $\l_i=\l(P_i)\;.$ The Darboux-Egoroff metrics on this space
are written in terms of the Schiffer $\Omega(P,Q)$ and Bergman $B(P,\bar{Q})$ kernels on a Riemann surface of genus $g\geq 1\;.$ These kernels are defined by \cite{Fay92}:
\beqs
\Omega(P,Q)=W(P,Q)-\pi\sum_{i,j=1}^g(\imb)_{ij}^{-1}\omega_i(P)\omega_j(Q) \;,
\eeqs
\beqs
B(P,\Qbar)=\pi\sum_{i,j=1}^g(\imb)_{ij}^{-1}\omega_i(P)\overline{\omega_j(Q)}\;,
\eeqs
where $W(P,Q)={d}_P{d}_Q\log E(P,Q)$ is the canonical bidifferential of the second kind on $\surf\;;$
$E(P,Q)$ is the prime form; $\{\omega_i\}_{i=1}^g$ are holomorphic differentials on $\surf$ normalized with respect to a given canonical basis of cycles by $\oint_{a_i}\omega_j=\delta_{ij} \;;$ and $\B$ is the symmetric matrix of their $b$-periods: $\B_{ij}=\oint_{b_i}\omega_j\;.$

The kernels can equivalently be characterized as follows \cite{Fay92}. 
The Schiffer kernel is the bidifferential with a singularity of the form $(x(P)-x(Q))^{-2}dx(P)dx(Q)$ along the diagonal $P=Q$ such that 
$ p.v. {\iint}_\surf \Omega(P,Q) \; \overline {\omega(P)} = 0 $
holds for any holomorphic differential $\omega$ on the surface. 
The Bergman kernel is a regular bidifferential on $\surf$ holomorphic with respect to its first argument and antiholomorphic with respect to the second one. It is (up to a factor of $2\pi i$) a kernel of an integral operator acting in the space $L_2^\h(\surf)$ of (1,0)-forms as an orthogonal projector onto the subspace ${\cal H}^\h(\surf)$ of holomorphic (1,0)-forms. In particular, for any holomorphic differential $\omega$ on the surface $\surf$ the following relation holds: $\iint_\surf B(P,\bar{Q}) \; \omega(Q) = 2\pi i \omega(P) \;.$
Both kernels, $\Omega(P,Q)$ and $B(P,\bar{Q})\;,$ are independent of the choice of a canonical basis of cycles $\{a_k,b_k\}\,.$ 

We consider the following family of metrics on the real Hurwitz space: 
\beqn
{\bf ds^2} = \sum_{j=1}^L \left(\oint_lh(Q) \Omega(Q,P_j) \right)^2 (d\l_j)^2 + \sum_{j=1}^L \left( \oint_lh(Q) B(Q,\bar{P_j}) \right)^2 (d\lb_j)^2 \;.
\label{metr}
\eeqn
Here $l$ is an arbitrary contour on the surface not passing through ramification points and such that its projection on the base of the covering does not depend on coordinates $\{\l_i;\lb_i\};$ $h$ is an arbitrary function defined in a neighbourhood of the contour. 
The rotation coefficients $\beta_{ij}$ of the metrics (\ref{metr}) are given by the Schiffer and Bergman kernels evaluated at the ramification points of the covering with respect to the local parameters given by $\sqrt{\l(P)-\l_i}\;:$
\beqs
\beta_{ij}=\Omega(P_i,P_j) \;, \qquad \beta_{i\bar{j}}=B(P_i,\bar{P}_j) \;, \qquad \beta_{\ib\jb}=\overline{\Omega(P_i,P_j)} \;.
\eeqs
As a consequence of Rauch variational formulas for the Schiffer and Bergman kernels, we have relations $\d_{\l_k} \beta_{ij} = \beta_{ik} \beta_{kj}$ for the rotation coefficients for distinct indices $i,j,k$ from the set $\{m;\bar{m}\}_{m=1}^\iL \;.$ These relations provide main conditions for the flatness of metrics (\ref{metr}). 

Some of the metrics (\ref{metr}) correspond to Frobenius structures on the Hurwitz space. We describe these structures and find their prepotentials and flat coordinates of the corresponding flat metric. A prepotential as a function of flat coordinates satisfies the WDVV system.

Since for the surface of genus zero the Bergman kernel vanishes and the Schiffer kernel coincides with $W(P,Q)\;,$ the metrics (\ref{metr}) and therefore the construction of  Frobenius ma\-ni\-folds suggested here is only new for a Hurwitz space in genus $\geq 1\;.$ For the Riemann sphere, our construction coincides with that of Dubrovin. 
For the simplest Hurwitz space in genus one, which has the real dimension $6,$ we compute explicitly prepotentials of three new Frobenius manifolds. One of these prepotentials has the form:
\begin{align}
\begin{split}
   F  & = -\frac{1}{4} t_1 t_2^2 - \frac{1}{4} t_1 t_5^2 + \frac{1}{2} t_1 t_4 (2 t_3 - \frac{1}{2 \pi i} ) - \frac{1}{2} t_1^2 t_6  - \frac{1}{2} t_3 (t_3 - \frac{1}{2 \pi i}) \frac{t_4^2}{t_6} - \frac{1}{16} \frac{t_2^2 t_5^2}{t_6}   \\
   & - 
   \frac{t_2^4}{32 t_6} - \frac{1}{128 \pi i} \frac{t_2^4}{t_6^{2}} \; \gamma \left( \frac{t_3}{t_6} \right) + \frac{t_3 t_4 t_5^2}{4 t_6}  \\
   & - 
   \frac{t_5^4}{32 t_6} - \frac{1}{128 \pi i} \frac{t_5^4}{t_6^{2}} \; \gamma \left( \frac{1 - 2 \pi i t_3}{2 \pi i t_6} \right) + \frac{(t_3 - \frac{1}{2 \pi i}) t_4 t_2^2}{4 t_6} \;,   
\end{split}
\label{Intro_prep}
\end{align}
where $\g(\mu)=4 \d_\mu \log\eta(\mu)$ for $\eta$ being the Dedekind $\eta$-function.
The function $F$ is quasihomogeneous, i.e. it satisfies 
\beqs
F(\kappa t_1,\kappa^{1/2}t_2,t_3, \kappa t_4,\kappa^{1/2}t_5,t_6) = \kappa^2 F(t_1,t_2,t_3,t_4,t_5,t_6)
\eeqs
for any nonzero constant $\kappa\;.$ The matrix $F_1$ formed by third derivatives $F_{t_1 t_i t_j}$ is constant and invertible; it gives the flat metric (written in flat coordinates) from the family of metrics (\ref{metr}) which corresponds to the Frobenius structure (\ref{Intro_prep}). 
The functions  
\beqs
c^k_{ij} = \sum_n (F^{-1}_1)_{kn}\frac{\d^3F}{\d t_i\d t_j\d t_n} 
\eeqs
define an associative commutative algebra in the tangent space to the underlying Hurwitz space: $\d_{t_i} \cdot \d_{t_j} = c^k_{ij}\d_{t_k}\;.$ (This is equivalent \cite{Dubrovin} to the WDVV system for the function $F$.)

Associated with any semisimple Frobenius manifold is the $G$-function, the solution to Getzler's system of linear differential equations derived in \cite{Getzler} within the study of recursion relations for the genus one Gromov-Witten invariants of smooth projective varieties. This system may be written for any semisimple Frobenius manifold. In \cite{DubZhang} it was proven that, for an arbitrary semisimple Frobenius manifold, the Getzler system has a unique quasihomogeneous solution given by 
\beqn
G = \log \frac{\tau_\iI}{J^{\scriptscriptstyle{1/24}}}  \;.
\label{Intro_G}
\eeqn
Here $J$ is the Jacobian of the transformation between canonical and  flat coordinates on the Frobenius manifold; $\tau_\iI$ is the isomonodromic tau-function associated to the Frobenius manifold. For the Frobenius structures described here the ingredients of the formula (\ref{Intro_G}) can be computed using results of papers \cite{KokKorB,KokKorG}. For example, the isomonodromic tau-function $\tau_\iI$ of the new Frobenius manifolds is related to the isomonodromic tau-function $\tau_\iI^\i0$ of Dubrovin's Hurwitz Frobenius manifolds by the formula:
\beqs
\tau_\iI = |\tau_\iI^\i0|^2 \left(\det {\rm Im}\B\right)^{-\frac{1}{2}}\;,
\eeqs
where $\B$ is the matrix of $b$-periods of the underlying Riemann surface. The function $\tau_\iI^{-2}$ coincides with an appropriately regularized ratio of the determinant of Laplacian on the Riemann surface and the surface volume in the singular metric $|d\l|^2\;,$ see \cite{ D'HokerPhong, KokKorB, Sonoda}.
\nopagebreak

For the Frobenius manifold corresponding to the prepotential (\ref{Intro_prep}), the G-function is expressed in terms of the Dedekind eta-function as follows: 
\beqs
G = - \log \left\{ \eta \left( \frac{t_3}{t_6} \right) \eta \left( \frac{1 - 2 \pi i t_3}{ 2 \pi i t_6} \right) \left( t_2 t_5 \right)^\frac{1}{8} t_6^{-\frac{3}{4}}  \right\} + {\rm const} \;.
\eeqs

We hope that in the future the construction of a ``real double" can be extended to arbitrary Frobenius manifolds. Presumably this extension can be done on the level of the Riemann-Hilbert problem associated with a Frobenius manifold. The most intriguing case would then be the Frobenius manifolds related to quantum cohomologies; we hope that their ``real doubles" might find an interesting geometrical application. 

We notice that a class of solutions to the WDVV system related to real Hurwitz spaces was previously constructed in the work \cite{Dzhamay}. However, the full structure of a Frobenius manifold was not discussed in \cite{Dzhamay}, and an explicit relationship of prepotentials of \cite{Dzhamay} and solutions to WDVV equations constructed in this work remains unclear. 

The paper is organized as follows. In the next section we give definitions of the WDVV system and Frobenius manifold and discuss the one-to-one correspondence between them. In Section \ref{Kernels} we describe the Hurwitz space we shall build Frobenius structures on, the $W$-bidifferential and the Schiffer and Bergman kernels on a Riemann surface and introduce flat metrics on Hurwitz spaces in terms of the kernels. In Section \ref{DubFrob} we reformulate the structures of Frobenius manifolds on Hurwitz spaces introduced by Dubrovin in terms of the $W$-bidifferential. Section \ref{NewFrob} contains the main result of the paper, the Frobenius structures on Hurwitz spaces considered as real manifolds. Section \ref{G-function} is devoted to calculation of the $G$-function for the new Frobenius structures. In Section \ref{Examples} we consider the simplest Hurwitz space in genus one and present explicit expressions for prepotentials and $G$-functions of the corresponding Frobenius manifolds. 

\section{Frobenius manifolds and WDVV equations}
\label{definitions}
The Witten - Dijkgraaf - E.Verlinde - H.Verlinde (WDVV) system looks as follows: 
\beqs
F_iF_1^{-1}F_j=F_jF_1^{-1}F_i\;,\hsp i,j=1,\dots,n \;,
\eeqs
where $F_i$ is the $n\times n$ matrix 
\beqs
(F_i)_{lm}=\frac{\d^3F}{\d t^i\d t^l\d t^m}\;,
\eeqs
and $F$ is a scalar function of $n$ variables $t^1,\dots,t^n\;.$ In the theory of Frobenius manifolds one imposes the following two conditions on the function $F\;:$
\begin{itemize}
\item Quasihomogeneity (up to a quadratic polynomial): for any nonzero 	$\kappa$ and some numbers $\nu_1,\dots,\nu_n, \nu_\iF$ 
	\beqn
	F(\kappa^{\nu_1}t^1,\dots,\kappa^{\nu_n}t^n)=\kappa^{\nu_\iF}F(t^1,\dots,t^n) + \mbox{quadratic terms} \;,
	\label{quasihomogeneity}
	\eeqn
\item Normalization: $F_1$ is a constant nondegenerate matrix.
\end{itemize}

The condition of quasihomogeneity can be rewritten in terms of the Euler vector field
\beqn
E := \sum_{\alpha}\nu_\alpha t^\alpha \d_{t^\alpha} \;
\label{Euler_flat}
\eeqn
as follows:
\beqn
\Lie_\iE F = E (F) = \sum_{\alpha} \nu_\alpha t^\alpha \d_{t^\alpha} F = \nu_\iF F + \mbox{quadratic terms}\;.
\label{Euler_quasihom}
\eeqn
\begin{definition}
An algebra $A$ over $\C$ is called a (commutative) {\bf Frobenius algebra} if: 
\itemize
\item it is a commutative associative $\C$-algebra with a unity ${\bf e}\;.$ 
\item it is supplied with a $\C$-bilinear symmetric nondegenerate inner product $\langle\cdot,\cdot\rangle$ having the property $\langle x\cdot y,z\rangle =\langle x,y\cdot z\rangle$ for arbitrary vectors $x,y,z$ from $A\;.$ 
\end{definition}
\begin{definition}
\label{Frobman}
$M$ is a {\bf Frobenius manifold} of the charge $\nu$ if a structure of a Frobenius algebra smoothly depending on the point $t\in M$ is specified on any tangent plane $T_tM$ such that
\enumerate
\item[{\bf F1}] the inner product $\langle\cdot ,\cdot\rangle $ is a flat metric on $M$ (not necessarily positive definite).
\item[{\bf F2}] the unit vector field ${\bf e}$ is covariantly constant with respect to the Levi-Civita connection $\nabla$ of the metric $\langle\cdot ,\cdot\rangle\;,$ i.e. $\nabla_{\bf x} {\bf e}=0$ for any vector field ${\bf x}$ on $M\;.$
\item[{\bf F3}] the tensor $(\nabla_{\bf w}{\bf c})({\bf x},{\bf y},{\bf z})$ is symmetric in four vector fields ${\bf x},{\bf y},{\bf z},{\bf w}\in T_tM\;,$ where ${\bf c}$ is the following symmetric $3$-tensor: ${\bf c}({\bf x},{\bf y},{\bf z})=\langle {\bf x}\cdot {\bf y},{\bf z}\rangle\;.$
\item[{\bf F4}] there exists on $M$ a vector field $E$ (the Euler field) such that the following conditions hold for any vector fields ${\bf x}\,,$ ${\bf y}$ on $M$
\beqn
\nabla_{\bf x}(\nabla_{\bf y} E)=0\;,
\label{covlin}
\eeqn
\beqn
[E,{\bf x}\cdot {\bf y}]-[E,{\bf x}]\cdot {\bf y} - {\bf x}\cdot[E,{\bf y}] = {\bf x}\cdot {\bf y}\;,
\label{Euler1}
\eeqn
\beqn
\Lie_\iE \langle {\bf x},{\bf y} \rangle := E \langle {\bf x},{\bf y} \rangle - \langle [E,{\bf x}],{\bf y} \rangle - \langle {\bf x},[E,{\bf y}] \rangle = (2-\nu) \langle {\bf x},{\bf y} \rangle \;.
\label{Euler2}
\eeqn
\end{definition}
The charge $\nu$ of a Frobenius manifold is equal to $\nu_\iF+3\;,$ where $\nu_\iF$ is the quasihomogeneity coefficient from (\ref{Euler_quasihom}).
\begin{theorem}(\cite{Dubrovin}) Any solution $F(t)$ of the WDVV equations with $\nu_1\neq 0$ defined for $t\in M$ determines on $M$ a structure of a Frobenius manifold and vice versa.
\label{thm1}
\end{theorem}

{\it Proof}
(see \cite{Dubrovin}). Given a Frobenius manifold, denote by $\{t^\alpha\}$ the flat coordinates of the metric $\langle \cdot,\cdot\rangle$ and by $\eta$ the constant matrix 
$\eta_{\alpha\beta}=\langle\d_{t^\alpha},\d_{t^\beta}\rangle\;.$
Due to the covariant constancy of the unit vector field ${\bf e}\;,$ we can by a linear change of coordinates put ${\bf e}=\d_{t^1}\;.$
In this coordinates, the condition {\bf F3} of Definition \ref{Frobman} implies the existence of a function $F$ whose third derivatives give the $3$-tensor ${\bf c}$:
\beqs
{\bf c}_{\alpha\beta\gamma}={\bf c}(\d_{t^\alpha},\d_{t^\beta},\d_{t^\gamma})=\frac{\d^3F}{\d t^\alpha\d t^\beta\d t^\gamma}\;.
\eeqs
The WDVV equations for the function $F$ provide the associativity condition for the Frobenius algebra defined by relations
$\d_{t^\alpha}\cdot\d_{t^\beta}={\bf c}^\gamma_{\alpha\beta}\d_{t^\gamma} \;,$
where the structure constants ${\bf c}^\gamma_{\alpha\beta}$ are found from
${\bf c}^\delta_{\alpha\beta}\eta_{\delta\gamma}={\bf c}_{\alpha\beta\gamma}\;.$
The existence of the vector field $E$ implies the quasihomogeneity of the function $F\;.$ Indeed, requirements (\ref{Euler1}), (\ref{Euler2}) on the Euler vector field imply  
\begin{multline}
\Lie_\iE {\bf c} (x,y,z) := E \left( {\bf c} (x,y,z) \right) - {\bf c} ([E,x],y,z) - {\bf c} (x,[E,y],z) - {\bf c} (x,y,[E,z]) \\
 = (3-\nu) {\bf c} (x,y,z) \;.
\label{Lie_c}
\end{multline}
The Lie derivative $\Lie_\iE$ commutes with the covariant derivative $\nabla$ as can easily be checked in flat coordinates when the Euler vector field (due to (\ref{covlin}))
has the form (\ref{Euler_flat}). Therefore, (\ref{Lie_c}) implies
$\Lie_\iE F = (3-\nu) F + \mbox{quadratic terms}\;. $

The converse statement can be proven analogously. 
$\Box$

The function $F\;,$ defined up to an addition of an arbitrary quadratic polynomial in $t^1,\dots,t^n\;,$ is called the {\it prepotential} of the Frobenius manifold.
\begin{definition}
A Frobenius manifold $M$ is called {\bf semisimple} if the Frobenius algebra in the tangent space at each point of $M$ does not have nilpotents.
\end{definition}

In this paper we only consider semisimple Frobenius structures.
\section{Kernels on Riemann surfaces and Darboux-Egoroff metrics}
\label{Kernels}
\subsection{Hurwitz spaces}
\label{Hurwitz}
Hurwitz space is the moduli space of pairs $( \surf,\lambda )$ where $\surf$ is a compact Riemann surface of genus $g$ and $\lambda:\surf\to\C P^1$ is a meromorphic function on $\surf$ of degree $N\;.$ The pair $(\surf,\l)$ represents the surface as an $N$-fold ramified covering $\cov$ of $\C P^1$ defined by the equation
\beqs
\zeta = \l(P) \;, \qquad P \in \surf 
\eeqs
($\zeta$ is a coordinate on $\C P^1$). In this way the surface $\surf$ can be viewed as a collection of $N$ copies of $\C P^1$ which are glued together along branch cuts. Critical points $P_j$ of the function $\l(P)$ correspond to ramification points of the covering. 
The projections $\l_j$ of ramification points on the base of the covering ($\C P^1$ with coordinate $\zeta$) are the images of critical points $P_j$ of the function $\l(P)$ ($\l_j$ are called the branch points):
$\l^\prime (P_j) = 0;\; \l_j = \l(P_j) \;.$

We assume that all finite branch points $\{ \l_j \; | \l_j < \infty \}$ are simple ( i.e. there are exactly two sheets glued together at the corresponding point) and denote their number by $L\;.$ We also assume that the function $\l$ has $m+1$ poles at the points of $\L$ denoted by $\infty^0,\dots,\infty^m;$ the pole at $\infty^i$ has the order $n_i+1\;.$ In terms of sheets of the covering, there are $m+1$ points which project to $\zeta=\infty$ on the base; the numbers $\{ n_i+1 \}$ give the number of sheets glued at each of these points ($n_0,\dots,n_m \in \N$ are such that  $\sum_{i=0}^m (n_i+1) = N\;,$ they are called the ramification indices).

The local parameter near a simple ramification point $P_j \in \L$ (which is not a pole of $\l$) is $x_j(P)= \sqrt{\l(P)-\l_j} \;;$ and in a neighbourhood $P\sim\infty^i$ the local parameter $z_i$ is given by $z_i(P) = \left( \l(P) \right)^{-1/(n_i+1)} \;.$

The Riemann-Hurwitz formula connects the genus $g$ of the surface, degree $N$ of the function $\l\;,$ the number  $L$ of simple finite branch points, and the ramification indices  $n_i$ over infinity:
\beqn
2 g - 2 = - 2N + L + \sum_{i=0}^m n_i \;.
\label{RH}
\eeqn

Two coverings are said to be equivalent if one can be obtained from the other by a permutation of sheets. 
The set of equivalence classes of described coverings will be denoted by $M=M_{g;n_0,\dots,n_m}\;.$
We shall work with a covering $\covM = \covM_{g;n_0,\dots,n_m}$ of this space. A point of the space $\covM$ is a triple $\{ \surf,\l,\{ a_k, b_k \}_{k=1}^g \}\,,$ where $\{ a_k, b_k \}_{k=1}^g$ is a canonical basis of cycles on $\surf\,.$ The branch points $\l_1,\dots,\l_\iL$ play the role of local coordinates on $\covM\;,$ viewed as a complex manifold. 
\subsection{Bidifferential $W\;,$ Bergman and Schiffer kernels}
First, we summarize properties of three well-known symmetric bidifferentials on Riemann surfaces. Being suitably evaluated at the ramification points $ \{ P_j \}\;,$ these kernels will play the role of rotation coefficients of flat metrics on Hurwitz spaces. 

The meromorphic bidifferential $W(P,Q)$ defined by
\beqn\label{W-def}
W(P,Q) := d_P d_Q\log E(P,Q)
\eeqn
is the symmetric differential on $\surf\times \surf$ with the second order pole at the diagonal $P=Q$ with biresidue $1$ and the properties:
\beqn
\oint_{a_k} W(P,Q) = 0 \;; \hsp \oint_{b_k} W(P,Q) = 2 \pi i \, \omega_k(P) \;; \hsp k=1, \dots, g \;.
\label{W-periods}
\eeqn
Here $\{a_k,b_k\}_{k=1}^g$ is the canonical basis of cycles on $\surf\;;\;\;$ $ \{ \omega_k(P) \}_{k=1}^g$ is the corresponding set of holomorphic differentials normalized by $\oint_{a_l}\omega_k=\delta_{kl}\;;$ and  $E(P,Q)$ is the prime form on the surface $\surf \;.$ The dependence of the bidifferential $W$ on branch points of the Riemann surface is given by the Rauch variational formulas \cite{KokKor, Rauch}:
\beqn
\frac{\d W(P,Q)}{\d\l_j}=\frac{1}{2}W(P,P_j)W(Q,P_j)\;,
\label{W-variation}
\eeqn
where $W(P,P_j)$ denotes the evaluation of the bidifferential $W(P,Q)$ at  $Q=P_j$ with respect to the standard local parameter $x_j(Q)=\sqrt{\l(Q)-\l_j}$ near the ramification point $P_j\;:$ 
\begin{equation}
W(P,P_j) := \frac{W(P,Q)}{dx_j(Q)} \vert_{Q=P_j} \;.
\label{notation}
\end{equation}
The bidifferential $W(P,Q)$ depends holomorphically on the branch points $\{\l_j\}$ in contrast to the following two bidifferentials \cite{Fay92}.

The {\it Schiffer kernel} $\Omega(P,Q)$ is the symmetric differential on $\surf \times \surf$ defined by:
\beqn
\Omega(P,Q) := W(P,Q)-\pi\sum_{k,l=1}^g(\imb)_{kl}^{-1}\omega_k(P)\omega_l(Q) \;,
\label{Omegadef}
\eeqn
where $\B$ is the symmetric matrix of $b$-periods of holomorphic normalized differentials $\{\omega_k\}\;:$ $\;\B_{kl}=\oint_{b_k}\omega_l\;,$ which depends holomorphically on the branch points $\{\l_j\} \;.$ This kernel has the same singularity structure as the bidifferential $W\;,$ it depends on $\{\lb_j\}$ due to the terms added to $W\;,$ since $\imb=(\B-\bar{\B})/(2i)$ and $\bar{\B}$ is a function of $\{\lb_j\}$. For a surface of genus zero the Schiffer kernel coincides with $W\;.$

The {\it Bergman kernel} $B(P,\Qbar)$ is defined by:
\beqn
B(P,\Qbar)=\pi\sum_{k,l=1}^g(\imb)_{kl}^{-1}\omega_k(P)\overline{\omega_l(Q)}\;.
\label{Bdef}
\eeqn
It vanishes for a surface of genus zero. 

An important property of the Schiffer and Bergman kernels is independence of the choice of a canonical  basis of cycles $ \{ a_k,b_k \}_{k=1}^g$ on the Riemann surface. 
This can be seen, for example, from the following definitions (see Fay \cite{Fay92}) equivalent to (\ref{Omegadef}) and (\ref{Bdef}). 

The Schiffer kernel is the unique symmetric bidifferential with a singularity of the form $(x(P)-x(Q))^{-2}dx(P)dx(Q)$ along $P=Q$ and such that 
\beqn
{\rm p.v.} \underset{\surf}{\iint} \Omega(P,Q) \overline {\omega(P)} = 0 
\eeqn
holds for any holomorphic differential $\omega \;.$ 

The Bergman kernel is (up to the multiplier $2\pi i$) a kernel of an integral operator which acts in the space $L_2^\h(\surf)$ of $(1,0)$-forms as an orthogonal projector onto the subspace ${\cal H}^\h(\surf)$ of holomorphic $(1,0)$-forms. 
In particular, the following holds for any holomorphic differential $\omega$ on the surface $\surf\;:$
\beqn
\frac{1}{2\pi i} \underset{\surf}{\iint} B(P,\bar{Q}) \omega(Q) = \omega(P)\;.
\label{Bdef2}
\eeqn
For the Bergman kernel the independence of the choice of a canonical basis of cycles can also be seen directly from (\ref{Bdef}) using 
$\;\left( {\rm Im} \B \right)_{kl} = \frac{i}{2} {\iint}_\surf \omega_k(P) \overline{\omega_l(P)}\;.\;$

The periods of Schiffer and Bergman kernels are related to each other as follows:
\beqn
\oint_{a_k} \Omega(P,Q) = - \oint_{a_k} B(\bar{P},Q) \;, \qquad \oint_{b_k} \Omega(P,Q) = - \oint_{b_k} B(\bar{P},Q)
\label{S-B_periods}
\eeqn
where the integrals are taken with respect to the first argument. Their derivatives with respect to branch points and their complex conjugates are given by:
\begin{equation}
\begin{split}
\frac{\d\Omega(P,Q)}{\d\l_j} = \frac{1}{2} \Omega(P,P_j) \Omega(Q,P_j) \;, 
\qquad
\frac{\d\Omega(P,Q)}{\d\bar{\l}_j} = \frac{1}{2} B(P,\bar{P}_j) B(Q,\bar{P}_j) \;, \\
\frac{\d B(P,\bar{Q})}{\d\l_j} = \frac{1}{2} \Omega(P,P_j) B(P_j,\bar{Q}) \;,
\qquad
\frac{\d B(P,\bar{Q})}{\d\bar{\l}_j} = \frac{1}{2} B(P,\bar{P}_j) \overline{\Omega(Q,P_j)} \;.
\label{SB-variation}
\end{split}
\end{equation}
The notation here is analogous to that in (\ref{notation}), i.e. $\Omega(P,P_j)$ stands for $\left( {\Omega(P,Q)} / {dx_j(Q)} \right) |_{Q=P_j}$ and $B(P,\bar{P}_j) := \left( B(P,\bar{Q}) / \overline{dx_j(Q)} \right) |_{Q=P_j} \;. $
To prove (\ref{SB-variation}) one uses the variational formulas (\ref{W-variation}) for $W(P,Q)\;,$ and the following Rauch variational foumulas for holomorphic normalized differentials $\{\omega_k\}$ and for the matrix of $b$-periods \cite{Rauch}:
\beqn
\frac{\d\,\omega_k(P)}{\d\l_j} = \frac{1}{2} \omega_k(P_j) W(P,P_j) \;, \qquad \frac{\d\,\B_{kl}}{\d\l_j} = \pi  i \, \omega_k(P_j) \omega_l(P_j) \;,
\label{Rauch}
\eeqn
where we write $\omega_k(P_j)$ for $(\omega_k(P) / dx_j(P))|_{P=P_j} \;.$ Derivatives of $\omega_k$ and $\B$ with respect to $\{ \lb_j \}$ vanish. 

\subsection{Darboux-Egoroff metrics}
Now we are in a position to introduce two families of Darboux-Egoroff (flat potential diagonal)  metrics on Hurwitz spaces written in terms of the described bidifferentials. Following the terminology of Dubrovin, we call a bilinear quadratic form a metric even if it is not positive definite. 

A diagonal metric 
${\bf ds^2}=\sum_i g_{ii}(d\l_i)^2$ is called potential if there exists a function $U$ such that $\d_{\l_i}U=g_{ii}$ for all $i\;.$ A potential diagonal metric is flat (Riemann curvature tensor vanishes) if its {\it rotation coefficients} $\beta_{ij}$ defined for $i \neq j$ by
\beqn
\beta_{ij} := \frac{\d_{\l_j}\sqrt{g_{ii}}}{\sqrt{g_{jj}}}
\label{rotation}
\eeqn
satisfy the system of equations:
\begin{align}
&\d_{\l_k} \beta_{ij} = \beta_{ik} \beta_{kj} \;,\hsp i,j,k\;\; \mbox{are distinct},
\label{flat1}
\\
&\sum_k \d_{\l_k} \beta_{ij} = 0 \qquad \mbox {for all} \;\;\beta_{ij}\;.
\label{flat2}
\end{align}

\subsubsection{Darboux-Egoroff metrics in terms of the bidifferential $W$}
The following family of diagonal metrics (bilinear quadratic forms) on the Hurwitz space first appeared in  \cite{KokKor} where it was realized that the corresponding rotation coefficients are given by the bidifferential $W$ (see (\ref{W-rotation})) and that the metrics are flat:
\beqn
{\bf ds^2} = \sum_{j=1}^L \left( \oint_l h(Q) W(Q,P_j) \right)^2 (d\l_j)^2 \;.
\label{metric1}
\eeqn
Here $l$ is an arbitrary smooth contour on the Riemann surface $\surf$ such that $P_j\notin l$ for any $j\;,$ and its image $\l(l)$ in $\C P^1$ is independent of the branch points $\{\l_j\};$ $h(Q)$ is an arbitrary independent of $\{\l_j\}$ function defined in a neighbourhood of the contour $l\;.$ 

Using variational formulas (\ref{W-variation}), we find that rotation coefficients of the metric (\ref{metric1}) are given by the bidifferential $W(P,Q)$ evaluated at the ramification points of the surface $\surf$ with respect to the standard local parameters $x_j = \sqrt{\l-\l_j}$ near $P_j\;:$
\beqn
\beta_{ij} = \frac{1}{2} W(P_i,P_j) \;,\hsp i,j=1,\dots,L\;,\;\;i\neq j \;.
\label{W-rotation}
\eeqn
Here $W(P_i,P_j)\;,$ similarly to (\ref{notation}),  stands for $\left( W(P,Q)/(dx_i(P) dx_j(Q)) \right)|_{P=P_i,Q=P_j}\;.$ Note that 
rotation coefficients $\beta_{ij}$ (\ref{W-rotation}) are symmetric with respect to indices, therefore the metrics (\ref{metric1}) are potential. The next proposition shows that they are Darboux-Egoroff metrics. 
\begin{proposition} \cite{KokKor}
\label{metric1-prop}
Rotation coefficients (\ref{W-rotation}) satisfy equations (\ref{flat1}), (\ref{flat2}) and therefore metrics (\ref{metric1}) are flat.
\end{proposition}

{\it Proof.}
Variational formulas (\ref{W-variation}) with $P=P_i\,,$ $Q=P_k\;,$ for different $i,j,k$ imply relations (\ref{flat1}) for rotation coefficients (\ref{W-rotation}). Equations (\ref{flat2}) hold for the coefficients due to the invariance of $W(P,Q)$ with respect to biholomorphic maps of the Riemann surface. Namely, consider the covering $\cov^{\!\!\delta}$ obtained from $\cov$ by a simultaneous $\delta$-shift $\l \to \l+\delta $ on all sheets. The surface $\surf$ is mapped  by this transformation to $\surf^\delta$ so that the point $P\in\surf$ goes to  $P^\delta \in \surf^\delta$ which belongs to the same sheet of the covering as $P$ and is such that $\l(P^\delta) = \l(P)+\delta\;.$  Denote by $W^\delta$ the bidifferential $W$ on the surface $\surf^\delta\;.$ Since the transformation $\l\to\l+\delta$ is biholomorphic, we have
$W^\delta(P^\delta,Q^\delta)=W(P,Q) \;.$
The same relation is true for $W(P,Q)/(dx_i(P)dx_j(Q))$ when points $P$ and $Q$ are in neighbourhoods of ramification points $P_i$ and $P_j\;,$ respectively:
\beqn
\frac{W^\delta(P^\delta,Q^\delta)}{dx^\delta_i(P^\delta)dx^\delta_j(Q^\delta)} = \frac{W(P,Q)}{dx_i(P)dx_j(Q)} \;.
\label{temp}
\eeqn
Note that $x_i(P) = \sqrt{\l(P)-\l_i}$ does not change under a simultaneous shift of all branch points and $\l\;.$ After the substitution $P=P_i\;,$  $Q=P_j$ in (\ref{temp}) the differentiation with respect to $\delta$ at $\delta=0$ gives the sum of derivatives with respect to branch points:
$\sum_{k=1}^\iL \d_{\l_k}W(P_i,P_j) = 0 \;.$
Thus, the rotation coefficients (\ref{W-rotation}) satisfy also (\ref{flat2}). 
Therefore the metrics (\ref{metric1}) are flat. 
$\Box$

\subsubsection{Darboux-Egoroff metrics in terms of Schiffer and Bergman kernels}
Now let us consider the Hurwitz space $M$ as a real manifold, i.e. a manifold with a set of local coordinates formed by the branch points and their complex conjugates. As an analogue of the family of metrics (\ref{metric1}) on the space of coverings $ M = M_{g;n_0,\dots,n_m}$ with the local coordinates $ \{ \l_1,\dots,\l_L; \lb_1,\dots,\lb_L \} $ we consider the following two families of metrics:
\beqn
{\bf ds}_1^{\bf 2} = \sum_{j=1}^L \left(\oint_lh(Q) \Omega(Q,P_j) \right)^2 (d\l_j)^2 + \sum_{j=1}^L \left( \oint_lh(Q) B(Q,\bar{P_j}) \right)^2 (d\lb_j)^2
\label{metric2}
\eeqn
and
\beqn
{\bf ds}_2^{\bf 2} =
{\rm Re} \left\{ 
 \sum_{j=1}^L \left( \oint_lh(Q) \Omega(Q,P_j) + \oint_l \overline{h(Q)}  B(\bar{Q},{P_j}) \right)^2 (d\l_j)^2 \right\} \;.
\label{metric2'}
\eeqn
Here, as before, $l$ is an arbitrary contour on the surface not passing through $\{P_j\}$ and such that its image $\l(l)$ in $\zeta$-plane is independent of branch points $\{\l_j\}\;;$ $h$ is an arbitrary function independent of $\{\l_j\}$  defined in some neighbourhood of the contour. 

From variational formulas (\ref{SB-variation}) for the Schiffer and Bergman kernels  we see that these metrics are potential and their rotation coefficients are given by the kernels evaluated at ramification points of $\surf \;:$
\beqn
\beta_{ij} = \frac{1}{2} \Omega(P_i,P_j) \;, \qquad
\beta_{i\jb} = \frac{1}{2} B(P_i,\bar{P_j}) \;, \qquad
\beta_{\ib\jb} = \overline{\beta_{ij}} \;.
\label{SB-rotation}
\eeqn
Here $i,j=1,\dots,L$ and the index $\jb$ corresponds to differentiation with respect to $\lb_j\;.$
Similarly to the notation in (\ref{W-rotation}), we understand $\Omega(P_i,P_j)$ and $B(P_i,\bar{P_j})$ as follows:
\beqs
\Omega(P_i,P_j) := \frac{\Omega(P,Q)}{dx_i(P) dx_j(Q)}{\Big|}_{P=P_i,\; Q=P_j}\;, \qquad 
B(P_i,\bar{P_j}) := \frac{B(P,\bar{Q})}{dx_i(P) \overline{dx_j(Q)}}{\Big|}_{P=P_i,\; Q=P_j}.
\eeqs
\begin{remark} \rm Note that rotation coefficients of the metrics (\ref{metric2}), (\ref{metric2'}) are defined on the space $M_{g;n_0,\dots,n_m}\;,$ in contrast to rotation coefficients (\ref{W-rotation}). The coefficients (\ref{W-rotation}) are given by the bidifferential $W\;,$ which depends on the choice of a canonical basis of cycles $\{a_i,b_i\}\;,$ and therefore are defined on the covering $\covM_{g;n_0,\dots,n_m}$ (see Section \ref{Hurwitz}). However, the metrics of the type (\ref{metric2}), (\ref{metric2'}) which will be used in Section \ref{NewFrob} still depend on the choice of cycles $ \{ a_i, b_i \} $ through the choice of contours $l\;.$ \end{remark}
\begin{proposition}
Rotation coefficients (\ref{SB-rotation}) satisfy equations (\ref{flat1}), (\ref{flat2}) and therefore metrics (\ref{metric2}), (\ref{metric2'}) are flat. 
\label{metric2-prop}
\end{proposition}

The proof is analogous to that of Proposition \ref{metric1-prop}. Here $\delta$ should be taken real, $ \delta \in \R \;.$

Note that in equations (\ref{flat1}), (\ref{flat2}) $i,j,k$ run through the set of all possible indices which in this case is $\{1,\dots,L;\bar{1},\dots,\bar{L}\}\;,$ where we put $\l_{\kb}:=\lb_k\;.$

\section{Dubrovin's Frobenius structures on Hurwitz spaces}
\label{DubFrob}

We start with a description of Dubrovin's construction \cite{Dubrovin} of Frobenius manifolds on the space $\covM = \covM_{g;n_0,\dots,n_m}$ using the bidifferential $W(P,Q)\;.$ The branch points $\l_1,\dots,\l_L$ are the local coordinates on $ \covM \;. $ 

To introduce a structure of a Frobenius algebra on the tangent space $T_t\covM$ for some point $t\in \covM$ we take coordinates ${\l_1,\dots,\l_L}$ to be canonical for multiplication, i.e we define
\beqn
\d_{\l_i} \cdot \d_{\l_j} = \delta_{ij} \d_{\l_i} \;.
\label{multiplication}
\eeqn
Then, the unit vector field is given by
\beqn 
{\bf e} = \sum_{i=1}^L \d_{\l_i} \;.
\label{e}
\eeqn

For this multiplication law, the diagonal metrics (\ref{metric1}) obviously have the property \linebreak $\langle {\bf x}\cdot {\bf y}, {\bf z}\rangle = \langle {\bf x},  {\bf y}\cdot {\bf z}\rangle$ required in the definition of a Frobenius algebra. Therefore together with the multiplication (\ref{multiplication}) the metrics (\ref{metric1}) define a family of Frobenius algebras on $T_t\covM\;.$

Among the family of metrics (\ref{metric1}) (and Frobenius algebras) we are going to isolate those corresponding to Frobenius manifolds.

The Euler vector field has the following form in canonical coordinates \cite{Dubrovin}:
\beqn
E = \sum_{i=1}^L \l_i \d_{\l_i} \;.
\label{Euler}
\eeqn

\subsection{Primary differentials}

As is easy to see, with the Euler field (\ref{Euler}), the multiplication (\ref{multiplication}) satisfies requirement (\ref{Euler1}) from {\bf F4}.
Condition (\ref{Euler2}) then reduces to
\beqn
E \left( \langle \d_{\l_i} , \d_{\l_i} \rangle \right) = -\nu \langle \d_{\l_i} , \d_{\l_i} \rangle \;.
\label{Eulermetric}
\eeqn
The following proposition describes the metrics from family (\ref{metric1}) which satisfy this condition. 
\begin{proposition}
\label{propEulermetric}
Let the contour $l$ in (\ref{metric1}) be either a closed contour on $\surf$ or a contour connecting points $\infty^i$ and $\infty^j$ for some $i$ and $j\;.$ In the latter case we regularize the integral by omitting its divergent part as a function of the corresponding local parameter near $\infty^i\;.$ Choose a function $h(Q)$ in (\ref{metric1}) to be $h(Q)=\const \,\l^n(Q)$ (where $\const$ is a constant). Then the Euler vector field (\ref{Euler}) acts on metrics (\ref{metric1}) according to (\ref{Eulermetric}) with $\nu = 1 - 2n \;.$
\end{proposition}

{\it Proof.}
Let us again use the invariance of the bidifferential $W$ under biholomorphic mappings of the Riemann surface $\surf \;.$ Consider the mapping $\cov \to \cov^{\!\!\epsilon}$ when the transformation $\l\to(1+\epsilon )\l$ is performed on every sheet of the covering $\cov\;.$ A point $P$ of the surface is then mapped to the point $P^\epsilon$ of the same sheet such that $\l(P^\epsilon)=(1+\epsilon)\l(P)\;.$  If  $W^\epsilon$ is the bidifferential $W$ on $\cov^{\!\!\epsilon}\;,$ then  
$W^\epsilon(P^\epsilon,Q^\epsilon)=W(P,Q)\;.$ For the local parameter $x_i=\sqrt{\l-\l_i}$ in a neighbourhood of a ramification point $P_i\;,$ we have $dx_i^\epsilon=\sqrt{1+\epsilon}\;dx_i\;.$
A contour $l$ of the specified type is invariant as a path of integration in (\ref{metric1}) with respect to this transformation. Therefore we have
\beqn
\left( \oint_{l^\epsilon} \l^n(Q^\epsilon) \frac{W^\epsilon(Q^\epsilon,P^\epsilon)}{dx^\epsilon_j(P^\epsilon)} \right)^2 = (1 + \epsilon)^{2n-1} \left( \oint_{l} \l^n(Q) \frac{W(Q,P)}{\;dx_j(P)} \right)^2 \;.
\label{temp1}
\eeqn
Putting $P=P_j \;,$ we differentiate (\ref{temp1}) with respect to $\epsilon$ at $\epsilon=0 \;.$ This yields the action of the vector field $E$ on the metric coefficient in the left-hand side  and proves the proposition. 
$\Box$
\begin{proposition}
Rotation coefficients (\ref{W-rotation}) given by the bidifferential $W$ satisfy $E \left( \beta_{ij} \right) = - \beta_{ij}\;.$
\end{proposition}

{\it Proof.} 
This is a corollary of Proposition \ref{propEulermetric} and can be proven by a straightforward calculation using (\ref{Eulermetric}) and the definition of rotation coefficients (\ref{rotation}). Alternatively, it can be proven directly by the method used in the proof of  Proposition \ref{propEulermetric}.
$\Box$

So far we have restricted the family of flat metrics to those of the form (\ref{metric1}) with $h = \const\,\l^n$ and the contour $l$ being either closed or connecting points $\infty^i\;,$ $\infty^j\;:$
\beqn
{\bf ds^2} = \sum_{j=1}^L \left( \const \oint_{l} \l^n(Q) W(Q,P_j) \right)^2 (d\l_j)^2 \;.
\label{rest_metr1}
\eeqn
An additional restriction comes from {\bf F2}, the requirement of covariant constancy of the unit vector field (\ref{e}) with respect to the Levi-Civita connection.
\begin{lemma}
If a diagonal metric ${\bf ds^2} = \sum_i g_{ii} (d\l_i)^2$ is potential (i.e. $\d_{\l_i}g_{jj} = \d_{\l_j}g_{ii}$ holds) and its coefficients $g_{ii}$ are annihilated by the unit vector field (\ref{e}) $ ( {\bf e}(g_{ii})=0 ) , $ then the vector field ${\bf e}$ is covariantly constant with respect to the Levi-Civita connection of the metric ${\bf ds^2}\;.$
\label{covarconst}
\end{lemma}
The proof is a simple calculation using the expression for the Christoffel symbols via coefficients of a diagonal metric:
\beqn
\Gamma_{ii}^k = -\frac{1}{2} \frac{\d_{\l_k}g_{ii}}{g_{kk}} \;, \qquad \Gamma_{ii}^i = \frac{1}{2} \frac{\d_{\l_i}g_{ii}}{g_{ii}} \;, \qquad \Gamma_{ij}^i = \frac{1}{2} \frac{\d_{\l_j}g_{ii}}{g_{ii}}\;, \qquad \Gamma_{ij}^k=0 \;\;\; \mbox{for distinct }  i, j, k .
\label{diagChristoffel}
\eeqn

Thus, we need to find the metrics of the form (\ref{rest_metr1}) such that the unit vector field ${\bf e}$ annihilates their coefficients. These metrics can be written as 
\beqn{\bf ds}_\phi^{\bf 2} = \sum_{i=1}^L \left( \underset{P=P_i}{\res} \; \frac{\phi^2(P)}{d\l(P)} \right) (d\l_i)^2 \equiv \frac{1}{2} \sum_{i=1}^L \phi^2(P_i) (d\l_i)^2 \;,
\label{phimetric}
\eeqn
where $\phi$ is a differential of one of the five types listed below in Theorem \ref{thm_primary}. These differentials are called primary and all have the form $\phi(P)= \const \oint_{l} \l^n(Q)W(Q,P)$ with some specific choice of a contour $l$ and function $\const \l^n\;.$ In other words, we shall consider five types of combinations of a contour and a function $\const\l^n\;.$ Let us write these combinations in the form of operations of integration over the contour with the weight function. The operations, applied to a $1$-form $f\;,$ have the following form:
\begin{alignat*}{2}
{\bf 1.\;\;} & \I_{t^{i;\a}} [f(Q)] := \frac{1}{\a} \; \underset{\infty^i} {\res} \;  \l(Q)^\frac{\a}{n_i+1} f(Q)
 \qquad &  i&=0, \dots, m\;; \; \a=1, \dots, n_i \;.\\
{\bf 2.\;\;} & \I_{v^i} [f(Q)] := 
\; \underset{\infty^i} {\res} \; \l(Q)f(Q) & \qquad   i&=1,\dots,m \;.\\
{\bf 3.\;\;} & \I_{w^i} [f(Q)] := \mathrm{v.p.} \int_{\infty^0}^{\infty^i} f(Q) & \qquad i&=1,\dots,m\;. \\
{\bf 4.\;\;} & \I_{r^k} [f(Q)] := - \oint_{a_k} \l(Q) f(Q) & \qquad k&=1, \dots, g\;.\\
{\bf 5.\;\;} & \I_{s^k} [f(Q)] := \frac{1}{2\pi i} \oint_{b_k} f(Q) & \qquad k&=1,\dots,g\;.
\end{alignat*}
Here the principal value near infinity is defined by omitting the divergent part of the integral as a function of the local parameter $z_i$  (such that $ \l=z_i^{-n_i-1})\;.$ 
\begin{theorem}
\label{thm_primary}
Let us choose a point $P_0 \in \surf$ which is mapped to zero by the function $\l\;,$ i.e. $\l(P_0)=0\;,$ and let all basis contours $\{a_k,b_k\}$ start at this point. Then, the defined operations 1.-5. applied to the bidifferential $W$ give a set of $L$ differentials, called primary,  with the following singularities (characteristic properties). 
By $z_i$ we denote the local parameter near $\infty^i$  such that $z_i^{-n_i-1}=\l\;,$ $n_i$ being the ramification index at $\infty^{i}\;.$ 
\begin{alignat*}{4}
&{\bf 1.}\;\; \phi_{t^{i;\a}}(P) &:=&\; \I_{t^{i;\a}}[W(P,Q)] && \hsp \sim z_i^{-\a-1}(P)dz_i(P) \;,  \; P \sim \infty^i\;; & \;\; i=0, & ... ,m\;; \; \a=1, ... , n_i\;. \\
&{\bf 2.}\;\;  \phi_{v^i}(P) &:=&\; \I_{v^i}[W(P,Q)] && \hsp \sim -d\l(P) \;, \;\; P \sim \infty^i \;; & \hsp  i&=1,\dots,m\;. \\
&{\bf 3.}\;\;  \phi_{w^i}(P) &:= &\; \I_{w^i}[W(P,Q)]: && \hsp \underset{\infty^i}{\res} \; \phi_{w^i} = 1 \;; \;\; \underset{\infty^0}{\res} \; \phi_{w^i} = -1 \;; & \hsp i&=1,\dots,m\;. \\
&{\bf 4.}\;\;   \phi_{r^k}(P) &:= &\; \I_{r^k}[W(P,Q)]: && \hsp \phi_{r^k}(P^{b_k}) - \phi_{r^k}(P) = 2 \pi i d\l(P) \;; & \hsp k&=1, \dots, g \;. \\
&{\bf 5.}\;\;   \phi_{s^k}(P) &:= &\; \I_{s^k}[W(P,Q)]: && \hsp \mbox{ holomorphic differential}
& \hsp k&=1,\dots,g \;. 
\end{alignat*}
Here $\phi_{r^k}(P^{b_k})-\phi_{r^k}(P)$ denotes the transformation of the differential under analytic continuation along the cycle $b_k$ on the Riemann surface. 

All above differentials have zero $a$-periods except $\phi_{s^l}$ which satisfy:  $\oint_{a_k}\phi_{s^l}=\delta_{kl}\;.$
\end{theorem}

{\it Proof.} Let us prove that 
\beqn
\phi_{t^{i;\a}} (P) \underset{P \sim \infty^i}{\sim} z_i^{-\a-1}(P)dz_i(P) \;.
\label{prooftemp}
\eeqn
It is easy to see that the differential $\phi_{t^{i;\a}} (P)$ has a singularity only at $P=\infty^i\;.$ Let us consider the expansion of the bidifferential $W$ at $Q\sim\infty^i:$
\beqn
W(P,Q) \underset{Q\sim\infty^i}{\simeq} W(P,\infty^i) + W^\prime_{,\;2}(P,\infty^i)z_i(Q) + \frac{1}{2}W^{\prime\prime}_{,\;2}(P,\infty^i) z_i^2(Q) + \dots .
\label{expan}
\eeqn
Since $W (P,Q)\simeq \left( (z_i(P) - z_i(Q) )^{-2} +{\cal O}(1) \right) d z_i(P) d z_i(Q) $ when $P \sim Q \sim \infty^i$ then we have for the $(\a-1)$-th coefficient of the expansion (\ref{expan})
\beqs
\frac{1}{\a !}W^{(\a-1)}_{,\;2}(P,\infty^i) \underset{P \sim \infty^i}{\sim} \frac{d z_i(P)}{z_i^{\a+1}(P)} \;,
\eeqs
which proves (\ref{prooftemp}). The case $\a = n_i + 1$ proves $\phi_{v^i}(P) \underset{P\sim\infty^i}{\sim} - d\l(P)\;.$

For the differentials $\phi_{\omega^i}$  the theorem can be proven analogously.

The differential $\phi_{r^k}(P)$ is not defined at the points of the contour $a_k\;,$ however it has certain limits as $P$ approaches the contour from different sides; thus $\phi_{r^k}(P)$ is defined and single valued on the fundamental polygon $\fpoly$ of the surface. (The fundamental polygon $\fpoly$ is obtained by cutting the surface along all basis cycles $a_k$ and $b_k$ provided they all start at one point.) Let us denote $d q_{k}^i (P) := \phi_{r^k}(P^{b_i}) - \phi_{r^k}(P) \;$
(as we shall see below, $d q_{k}^i$ is indeed an exact differential) and consider the differential $\phi_{r^k}(P) \int_{P_0}^P \omega_k\;$ ($\omega_k$ is one of the normalized holomorphic differentials such that $\oint_{a_j} \omega_k = \delta_{jk}$). This differential has no poles inside $\fpoly\;.$ Therefore its integral over the boundary of $\fpoly$ equals zero. On the other hand, since the boundary $\d\fpoly$ consists of cycles $\{a_j \}$ and $\{ b_j \}$ the integral can be rewritten via periods of the differentials as follows:
\beqn
0 = \oint_{\d\tilde{\L}} \phi_{r^k}(P) \int_{P_0}^P \omega_k = \oint_{b_k}\phi_{r^k}  - \sum_j \oint_{a_j}\phi_{r^k} \B_{jk} + \sum_j \oint_{a_j} q_{k}^j \omega_k \;
\label{prooftemp1}
\eeqn
($\B_{jk} = \oint_{b_j} \omega_k$). Due to the choice of the point $P_0$ where all basis cycles start, we can change the order of integration in expressions $\oint_{b_k} \oint_{a_k} \l(Q)W(P,Q)$
as can be checked by a local (near the point $P_0$) calculation of the integral. Therefore we have
\beqs
\oint_{a_j}\phi_{r^k}(P) = 0 \;\; \mbox { for all } \; j \qquad \mbox {and} \qquad \oint_{b_k}\phi_{r^k}(P) = - 2 \pi i \oint_{a_k} \l(Q) \omega_k (Q) \;.
\eeqs
Then, the relation (\ref{prooftemp1}) takes the form
%
\beqs
0 = - 2 \pi i \oint_{a_k} \l(Q) \omega_k(Q) + \sum_j \oint_{a_j} q_{k}^j(Q) \omega_k(Q) \;,
\eeqs
and we conclude that $q_{k}^j(Q) = 2 \pi i \l(Q) \delta_{jk} \;.$

For differentials $\phi_{s^k}$ the statement of the theorem follows from properties (\ref{W-periods}) of the bidifferential $W\;.$ 
For all primary differentials (except $\phi_{s^k}$) $a$-periods are zero since they are zero for $W\;.$
$\Box$

\subsection{Flat coordinates}

For a flat metric there exists a set of coordinates in which coefficients of the metric are constant. These coordinates are called the {\it flat coordinates} of the metric.
In flat coordinates the Christoffel symbols vanish and the covariant derivative  $\nabla_{t^\iA} $ is the usual partial derivative $ \d_{t^\iA}\;.$ Therefore flat coordinates can be found from the equation $\nabla_{\bf x}\nabla_{\bf y} t = 0$ (${\bf x}$ and ${\bf y}$ are arbitrary vector fields on the manifold). In canonical coordinates this equation has the form:
\beqn
\d_{\l_i}\d_{\l_j}t=\sum_k\Gamma_{ij}^k\d_{\l_k}t\;,
\label{flat}
\eeqn
where the Christoffel symbols are given by (\ref{diagChristoffel}). For different $i,j,k\;,$ the Christoffel symbols of the metrics ${\bf ds}_\phi^{\bf 2}$ (\ref{phimetric}) have the form:
\beqs
\Gamma_{ii}^k = - \beta_{ik} \frac{\phi(P_i)}{\phi(P_k)} \;, \qquad
\Gamma_{ii}^i = - \sum_{j,\;j \neq i} \Gamma_{ij}^i \;, \qquad
\Gamma_{ij}^i = \beta_{ij} \frac{\phi(P_j)}{\phi(P_i)} \;, \qquad
\Gamma_{ij}^k =0 \;.
\eeqs
\begin{theorem}(\cite{Dubrovin})
The following functions give a set of flat coordinates of the metric ${\bf ds}_\phi^{\bf 2}$ (\ref{phimetric}):
\begin{alignat*}{4}
& t^{i;\a} &=& -(n_i+1) \I_{t^{i;1+n_i-\a}} [\phi] \qquad     
&& i = 0, \dots, m \;; \; \a = 1, \dots, n_i \\
& v^i &=& - \I_{w^i} [\phi]
&& i  = 1, \dots, m \\
& w^i &=& - \I_{v^i} [\phi]
&& i  = 1, \dots, m \\
& r^k &=&\; \I_{s^k} [\phi]
&& k  = 1, \dots, g \\
& s^k &=&\; \I_{r^k} [\phi]
&& k  = 1, \dots, g \;.
\end{alignat*}

Non-zero entries of the constant matrix of the metric in these coordinates are:
\begin{align*}
{\bf ds}_\phi^{\bf 2} ( \d_{t^{i;\a}}, \d_{t^{j;\b}} ) & = \frac{1}{n_i+1} \delta_{ij} \delta_{\a+\b,n_i+1} \;, \\
{\bf ds}_\phi^{\bf 2} ( \d_{v^i}, \d_{w^j} ) & = \delta_{ij} \;, \\
{\bf ds}_\phi^{\bf 2} ( \d_{r^k}, \d_{s^l} ) & = - \delta_{kl} \;.
\end{align*}
\end{theorem}
For notational convenience we denote an arbitrary flat coordinate by $t^\iA\;,$ and a primary differential by $\phi_{t^\iA} \;,$  i.e.
\beqs t^\iA \in \{ t^{i;\a} \;; \; v^i \;, w^i \;; \; r^k \;, s^k \; | i = 0, \dots, m \;; \; \a = 1, \dots, n_i \;; k = 1, \dots,g \} \;.
\eeqs
\begin{proposition}
\label{Euler_in_flat}
In flat coordinates $\{ t^\iA \}$ of the metric ${\bf ds}_\phi^{\bf 2}\;,$ the Euler vector field (\ref{Euler}) has the form (\ref{Euler_flat}) with coefficients $\{ \nu_\iA \}$ depending on the choice of a primary differential $\phi \;:$
\begin{itemize}
\item if $\phi = \phi_{t^{i_o;\a}}$ then
  	\begin{align*}
	E = \sum_{i=0}^m \sum_{\a=1}^{n_i} \left( 1+ \frac{\a}{n_{i_o}+1} \right.&- \left.\frac{\a}{n_i+1} \right) t^{i;\a} \d_{t^{i;\a}} + \sum_{i=1}^m \left( \frac{\a}{n_{i_o}+1} v^i \d_{v^i} + (1+\frac{\a}{n_{i_o}+1}) \omega^i \d_{\omega^i} \right) \\
	 &+ \sum_{k=1}^g \left( \frac{\a}{n_{i_o}+1} r^k \d_{r^k} + (1+\frac{\a}{n_{i_o}+1}) s^k 	\d_{s^k}\right) \;,
	\end{align*}
\item if $\phi = \phi_{v^{i_o}}$ or $\phi = \phi_{r^{k_o}}$ then
  	\beqs
	E = \sum_{i=0}^m \sum_{\a=1}^{n_i}(2 - \frac{\a}{n_i+1}) t^{i;\a} \d_{t^{i;\a}} + 	\sum_{i=1}^m \left( v^i \d_{v^i} + 2 \omega^i \d_{\omega^i} \right) 
	 + \sum_{k=1}^g \left( r^k \d_{r^k} + 2 s^k \d_{s^k} \right) \;,
	\eeqs
\item if $\phi = \phi_{\omega^{i_o}}$ or $\phi = \phi_{s^{k_o}}$ then
  	\beqs
	E = \sum_{i=0}^m \sum_{\a=1}^{n_i}(1 - \frac{\a}{n_i+1}) t^{i;\a} \d_{t^{i;\a}} + 	\sum_{i=1}^m  \omega^i \d_{\omega^i}
	 + \sum_{k=1}^g  s^k \d_{s^k} \;.
	\eeqs
\end{itemize}
\end{proposition}
\begin{proposition}(see \cite{Dubrovin})
\label{unityprop}
The unit vector field ${\bf e}$ (\ref{e}) in the flat coordinates of the metric ${\bf ds}_{\phi_{t^{A_0}}}^{\bf 2}$ has the form: ${\bf e}=-\d_{t^{A_0}}\;.$ 
\end{proposition}

Thus, the coordinate $t^{\iA_0}$ is naturally marked. Let us denote it by $t^{\scriptscriptstyle{1}}$ so that ${\bf e}=-\d_{t^1}\;.$

In flat coordinates the Christoffel symbols of the Levi-Civita connection vanish. Therefore the proposition implies that the unit vector field is covariantly constant ({\bf F2}).

\subsection{Prepotentials of Frobenius structures}

\begin{definition}
\label{def_prepotential}
A {\bf prepotential} of a Frobenius manifold is a function $F$ of flat coordinates of the corresponding metric such that its third derivatives are given by the symmetric $3$-tensor ${\bf c}$ from the definition of a Frobenius manifold ({\bf F3}):
\beqn
\frac{\d^3 F(t)}{\d_{t^A}\d_{t^B}\d_{t^C}}={\bf c}(\d_{t^A},\d_{t^B},\d_{t^C})={\bf ds}_{\phi}^{\bf 2}(\d_{t^A}\cdot\d_{t^B},\d_{t^C})\;.
\label{prepotentialdef}
\eeqn
\end{definition}
By presenting this function (defined up to a quadratic polynomial in flat coordinates) for each metric $\;{\bf ds}_\phi^{\bf 2}\;$ we shall prove the symmetry in four indices $\;({\scriptstyle A,\;B,\;C,\;D})\;$ of the tensor $(\nabla_{\d_{t^D}}{\bf c})(\d_{t^A},\d_{t^B},\d_{t^C})$ and therefore complete the construction of the Frobenius manifold. 

We shall denote the Frobenius manifold corresponding to the metric ${\bf ds}_\phi^{\bf 2}$ by $ \covM^\phi = \covM^\phi_{g;n_0,\dots,n_m}$ and its prepotential by $F_\phi \;.$
\begin{remark}\rm Proposition \ref{unityprop} implies that the third order derivatives (\ref{prepotentialdef}) are constant if one of the derivatives is taken with respect to the coordinate $t^1\;$ 
\beqs
\frac{\d^3F}{\d t^{\scriptscriptstyle{1}}\d t^\iA\d t^\iB}=-{\bf ds}_{
\phi_{t^{\scriptscriptstyle{1}}}}^{\bf 2}(\d_{t^A},\d_{t^B})\;.
\eeqs
\end{remark}
Before writing a formula for the prepotential we shall define a pairing of differentials.
Let $\omega^{(1)}$ and $\omega^{(2)}$ be two differentials on the surface $\surf$ holomorphic outside of the points $\infty^0,\dots,\infty^m$ with the following behaviour at $\infty^i\;:$
\beqn
\omega^{(\a)} = \sum_{n=-n^{(\a)}}^\infty c_{n,i}^{(\a)} z_i^n dz_i + \frac{1}{n_i+1} d \left( \sum_{n>0} r_{n,i}^{(\a)} \l^n \log \l \right) \;, \;\; P \sim \infty^i \;,
\label{inf_exp}
\eeqn
where $n^{(\a)}\in\Z$ and $c_{n,i}^{(\a)}\;,\;r_{n,i}^{(\a)}$ are some coefficients;  $z_i=z_i(P)$ is a local parameter near $\infty^i\;.$
Denote also for $k=1,\dots,g\;:$
\beqn
\oint_{a_k}\omega^{(\a)}=A_k^{(\a)}\;,
\label{a_period}
\eeqn
\begin{alignat}{3}
&\omega^{(\a)} (P^{a_k}) - \omega^{(\a)} (P) &= dp_k^{(\a)}(\l(P))\;,\hsp &p_k^{(\a)}(\l) &= \sum_{s>0} p^{(\a)}_{sk} \l^s \;,
\label{a_twist}\\
&\omega^{(\a)}(P^{b_k}) - \omega^{(\a)}(P) &= dq_k^{(\a)}(\l(P)) \;, \hsp &q_k^{(\a)}(\l) &= \sum_{s>0}q^{(\a)}_{sk} \l^s \;.
\label{b_twist}
\end{alignat}
Here, as before, $\omega(P^{a_k})$ and $\omega(P^{b_k})$ denote the analytic continuation of $\omega(P)$ along the corresponding cycle on the Riemann surface.

Note that if $\omega^{(\alpha)}$ is one of the primary differentials (defined in Theorem \ref{thm_primary}), then the coefficients $c_{n,i},$  $r_{n,i},$ $p_{sk},$ $q_{sk}$ and $A_k$ do not depend on coordinates.
\begin{definition}
For two differentials  whose singularity structures are given by (\ref{inf_exp}) - (\ref{b_twist}) define a {\bf pairing} $\F[\;,\;]$ as follows:
\begin{align*}
\begin{split}
\F[\omega^{(\a)}\;,\;\omega^{(\b)}] = \sum_{i=0}^m \left(\sum_{n\geq 0}\frac{c^{(\a)}_{-n-2,i}}{n+1}c^{(\b)}_{n,i} + c_{-1,i}^{(\a)} \mathrm{v.p.} \int_{P_0}^{\infty^i} \omega^{(\b)} - \mathrm{v.p.} \int_{P_0}^{\infty^i} \sum_{n>0}r_{n,i}^{(\a)} \l^n \omega^{(\b)} \right)\\
 + \frac{1}{2\pi i} \sum_{k=1}^g \left( - \oint_{a_k} q_k^{(\a)}(\l) \omega^{(\b)} + \oint_{b_k} p_k^{(\a)}(\l) \omega^{(\b)} + A_k^{(\a)} \oint_{b_ k} \omega^{(\b)} \right)\;,
\end{split}
\end{align*}
where $P_0$ is a marked point on the surface such that $\l(P_0)=0 \;.$  
\end{definition}
For any primary differential $\phi$ we consider a (multivalued on $\surf$) function $\p \;:$  
\beqn
\p(P) = \mathrm{v.p.} \int_{\infty^0}^P \phi \;.
\label{p-mult}
\eeqn
One can see that singularities of the differential $\p d\l$ can be described by formulas similar to (\ref{inf_exp}) - (\ref{b_twist}). The corresponding coefficients $c_{n,i},$  $r_{n,i},$ $p_{sk},$ $q_{sk}$ and $A_k$ for $\omega=\p d\l$ depend on coordinates $\{ \l_k \}$ in contrast to those for primary differentials. 
\begin{theorem} (\cite{Dubrovin}) The following function gives a prepotential of the Frobenius manifold $\covM^\phi \;:$
\beqn
 F_\phi = \frac{1}{2} \F [\p d\l\;,\; \p d\l] \;,
\label{prepotential}
\eeqn
where $p$ is the multivalued function (\ref{p-mult}). The third derivatives of $F_\phi$ are given by
\begin{equation}
\begin{split}
 \frac{\d^3 F_\phi (t)}{ \d_{t^A} \d_{t^B} \d_{t^C} } = {\bf c}(\d_{t^A},\d_{t^B},\d_{t^C}) = -\sum_{i=1}^L \; \underset{P_i}{\rm  res} \; \frac{\phi_{t^\iA} \phi_{t^\iB} \phi_{t^\iC}}{d\p d\l} \\
 \equiv - \frac{1}{2} \sum_{i=1}^L \frac{\phi_{t^\iA}(P_i) \phi_{t^\iB}(P_i) \phi_{t^\iC}(P_i)}{\phi(P_i)} \;.
\label{Dubrc_flat}
\end{split}
\end{equation}
\end{theorem}
\begin{theorem} (\cite{Dubrovin}) The second derivatives of the prepotential $F_\phi$ are given by the pairing of the corresponding primary differentials:
\beqs
\d_{t^A} \d_{t^B} F_\phi = \F[ \phi_{t^A}\;,\;\phi_{t^B}]\;.
\eeqs
\end{theorem}

For the described Frobenius manifold $\covM^\phi\;,$ the prepotential (\ref{prepotential}) is a quasihomogenous function of flat coordinates $\{ t^\iA \}$ of the metric ${\bf ds}_\phi^{\bf 2}\;,$ i.e. the following holds for some numbers $\{\nu_\iA\}$ and $\nu_\iF$ and any non-zero constant $\kappa \;:$
\beqs
F_\phi (\kappa^{\nu_1}t^1, \dots, \kappa^{\nu_n}t^n)=\kappa^{\nu_\iF} F_\phi (t^1,\dots,t^n) + {\mbox {quadratic terms}} \;.
\eeqs
This follows from the existence of the Euler vector field satisfying (\ref{covlin}) - (\ref{Euler2}) (see the proof of Theorem \ref{thm1}).

The coefficients of quasihomogeneity $\{ \nu_\iA \}$ are coefficients of the Euler vector field written in flat coordinates (see (\ref{quasihomogeneity}) - (\ref{Euler_quasihom})); they are given by Proposition \ref{Euler_in_flat}. The coefficient $\nu_\iF = 3 - \nu$ can be computed  for each Frobenius structure $\covM^\phi$ using Proposition \ref{propEulermetric}:
\begin{alignat*}{3}
& \; {\rm if} \;\;  \phi = \phi_{t^{i;\a}} \;, \;\;\;\;{\rm  then} & \nu = 1 - \frac{2\a}{n_i+1} &\qquad \nu_\iF = \frac{2\a}{n_i+1} +2 \\
&  \; {\rm if } \;\; \phi = \phi_{v^i} \;\;{\rm  or}\;\; \phi = \phi_{r^k}\;, \;\;{\rm  then} &  \nu = -1 & \qquad \nu_\iF = 4 \\
&  \; {\rm if } \;\; \phi = \phi_{\omega^i}\;\;{\rm  or} \;\; \phi = \phi_{s^k}\;, \;\;{\rm  then} & \nu = 1 & \qquad \nu_\iF = 2 \;.
\end{alignat*}
\begin{remark}\rm A linear combination of primary differentials corresponding to the same charge $\nu$ also gives a Frobenius structure. Namely, the above construction works for 
\beqs
\phi = \sum_{i=1}^m \kappa_i\phi_{v^i} + \sum_{k=1}^g \sigma_k \phi_{r^k} \qquad \mbox{and} \qquad
\phi = \sum_{i=1}^m \kappa_i\phi_{\omega^i} + \sum_{k=1}^g \sigma_k \phi_{s^k} \;,
\eeqs
with any constants $\{\kappa_i \}$ and $\{\sigma_k\} \;.$ 
The unit vector field in these cases, respectively, is given by
\beqs
{\bf e} = -\left( \sum_{i=1}^m \kappa_i\d_{v^i} + \sum_{k=1}^g \sigma_k \d_{r^k} \right) \qquad \mbox{and} \qquad
{\bf e} = - \left( \sum_{i=1}^m \kappa_i\d_{\omega^i} + \sum_{k=1}^g \sigma_k \d_{s^k} \right) \;.
\eeqs
After a linear change of variables, the unit field can be written as ${\bf e} = - \d_{\xi^1}$ for a new variable $\xi^1\;,$ since the coordinates $\{v^i\}$ and $\{r^k\}$ ($\{\omega^i\}$ and $\{s^k\}$)
have equal quasihomogeneity coefficients.
\end{remark}

\section[``Real doubles" of Dubrovin's Frobenius structures on Hurwitz spaces]{``Real doubles" of Dubrovin's Frobenius structures on \\ Hurwitz spaces}
\label{NewFrob}
In this section we consider the moduli space $\covM = \covM_{g;n_0,\dots,n_m}$ as a real manifold. The set of local coordinates is given by the set of branch points of the covering $\cov$ and their complex conjugates: 
$ \{ \l_1, \dots, \l_\iL \;; \; \lb_1, \dots, \lb_\iL \} \;.$ On the space $\covM$ with coordinates $\{\l_i;\lb_i\}$ we shall build a Frobenius structure in a way analogous to the one described in Section \ref{DubFrob}. The construction will be based on a family of flat metrics on $\covM (\{ \l_i;\lb_i \})$ of the type (\ref{metric2}), (\ref{metric2'}) with rotation coefficients given by the Schiffer and Bergman kernels. Since in genus zero the Schiffer kernel coincides with the bidifferential $W$ and the Bergman kernel vanishes, we only get essentially new metrics (and therefore new Frobenius structures) for Hurwitz spaces in genus greater than zero.

We start with a description of a Frobenius algebra in the tangent space. The coordinates $ \{ \l_1, \dots, \l_\iL \;; \; \lb_1, \dots, \lb_\iL \}$ are taken to be canonical for multiplication:
\beqn
\d_{\l_i}\cdot\d_{\l_j}=\delta_{ij}\d_{\l_i}\;,
\label{multiplicationR}
\eeqn 
where indices $i,j$ range now in the set of all indices, i.e. $i,j\in\set\;,$ and we put $\l_{\ib}:=\lb_i\;.$ The unit vector field of the algebra is given by 
\beqn
{\bf e}=\sum_{i=1}^L\left(\d_{\l_i}+\d_{\lb_i}\right).
\label{e2}
\eeqn
The role of an inner product of the Frobenius algebra is played by one of the metrics (\ref{metric2}), (\ref{metric2'}).
The new vector field $E\;,$ analogously, is
\beqn
E := \sum_{i=1}^L \left( \l_i \d_{\l_i} + \lb_i \d_{\lb_i} \right) \;.
\label{Eulerfield2}
\eeqn

\subsection{Primary differentials}

Together with the multiplication (\ref{multiplicationR}), the Euler field (\ref{Eulerfield2}) satisfies relation (\ref{Euler1}) of {\bf F4}. Its action (\ref{Euler2}) on a diagonal metric  takes the form:
\beqn
E \left( \langle \d_{\l_k} , \d_{\l_k} \rangle \right) = - \nu \langle \d_{\l_k} , \d_{\l_k} \rangle \;, \hsp k \in \set \;.
\label{Eulermetric2}
\eeqn
Among the metrics (\ref{metric2}), (\ref{metric2'}) we choose, similarly to Proposition \ref{propEulermetric}, those for which this condition holds.
\begin{proposition}
\label{propEulermetric2}
Let the contour $l$ in (\ref{metric2}), (\ref{metric2'}) be either closed or connecting points $\infty^i$ and $\infty^j$ for some $i,\;j\;.$ In the latter case we regularize the integral by omitting its divergent part as a function of the local parameter $z_i$ (or as a function of $\bar{z_i}$) near $\infty^i \;.$ 
Then the metrics (\ref{metric2}), (\ref{metric2'}) with $h(Q)=\const\l^n(Q)$ (where $\const$ is a constant)  satisfy (\ref{Eulermetric2}) with $\nu= 1 - 2n$ and the Euler field (\ref{Eulerfield2}).
\end{proposition}

{\it Proof.} The proof is the same as for Proposition \ref{propEulermetric}: we use the fact that Bergman and Schiffer kernels are invariant under biholomorphic mappings of the Riemann surface. The biholomorphic map to be taken in this case is $\l\to(1+\epsilon)\l\;,$ where $\epsilon$ is real. $\Box$
\begin{proposition} Rotation coefficients (\ref{SB-rotation}) given by the Schiffer and Bergman kernels satisfy $E \left( \beta_{ij} \right) = - \beta_{ij} \;, \; i,j \in \set \;,$
where the Euler field $E$ is given by (\ref{Eulerfield2}).
\end{proposition}

{\it Proof.} This statement is a corollary of Proposition \ref{propEulermetric2}; it can also be proven directly by using the invariance of the kernels under the mapping of Riemann surfaces $\cov \to \cov^{\!\!\epsilon}\;,$ $\l \to (1+\epsilon)\l \;,$ for $\epsilon \in \R \;.$ $\Box$

Among the metrics ${\bf ds^2}=\sum_i \left( g_{ii}(d\l_i)^2 + g_{\bar{i}\bar{i}}(d\lb_i)^2 \right)$ of the form (\ref{metric2}), (\ref{metric2'}) with $h=\const\l^n$ and a contour $l$ of the type required in Proposition \ref{propEulermetric2} only those ones correspond to Frobenius manifolds whose coefficients satisfy $ {\bf e} ( g_{ii} ) =  {\bf {e}} ( g_{\bar{i}\bar{i}} ) = 0$ ( ${\bf {e}}$ is the unit vector field (\ref{e2})). This follows from {\bf F2} and  Lemma \ref{covarconst}, which is obviously valid for the unit vector field (\ref{e2}) and diagonal potential metrics (\ref{metric2}), (\ref{metric2'}). Therefore we need to find the combinations of a contour $l$ and a function $h = \const\l^n$ such that formulas (\ref{metric2}), (\ref{metric2'}) give metrics whose coefficients are annihilated by the vector field ${\bf e}\;.$
We list those combinations in the form of operations $\I[f(Q)] = \oint_l\const\l^nf(Q)$ applied to a differential $f$ of the form $f=f_{\h}+f_{\ah}\;.$ We say that a differential is of the $(1,0)$-type if in a local coordinate $z$ it can be represented as $f_\h=f_1(z)dz\;,$ and is of the $(0,1)$-type
if in a local coordinate it has a form $f_\ah=f_2(\bar{z})d\bar{z}\;.$ We shall also call $f_\h$ and $f_\ah$ the holomorphic and antiholomorphic parts of a differential $f\;,$ respectively. 
We denote by $\tilde{\res}$ the coefficient in front of $d\bar{z}/\bar{z}$ in the Laurent expansion of a differential. As before, $z_i$ is the local parameter in a neighbourhood of $\infty^{i}$  such that $ z_i^{-n_i-1}(Q) = \l(Q) \;,\; Q \sim \infty^i \;.$\\
For $i=0,\dots,m;\;\a=1,\dots,n_i$ we define: 
\begin{alignat*}{2}
 {\bf 1.} \;\; \I_{t^{i;\a}} [f(Q)] & := 
\frac{1}{\a} \; \underset{\infty^{i}}{\res}  \;  z_i^{-\a}(Q) f_{\h}(Q)  \qquad
& {\bf 2.} \;\; \I_{t^{\overline{i;\a}}} [f(Q)] & := 
 \frac{1}{\a} \; \underset{\infty^{i} }{\tilde{\res}} \; \bar{z}_i^{-\a}(Q) f_{\ah}(Q)  \\
 {\bf 3.} \; \I_{v^i} [f(Q)]  &:= 
\; \underset{\infty^{i}}{\res} \; \l(Q) f_{\h}(Q) 
& {\bf 4.} \; \I_{v^{
\bar{i}}} [f(Q)] & := 
\; \underset{\infty^{i}}{\tilde{\res}} \; \lb(Q) f_{\ah}(Q) \;. 
\end{alignat*}
For $ i=1,\dots,m$ we define: 
\begin{alignat*}{2}
{\bf 5.} \; \I_{w^i} [f(Q)] & := \mathrm{v.p.} \int_{\infty^0}^{\infty^i} f_{\h}(Q) & \qquad {\bf 6.} \; \I_{w^{\bar{i}}} [f(Q)] & := \mathrm{v.p.} \int_{\infty^0}^{\infty^i} f_{\ah}(Q) \;.
\end{alignat*}
As before, the principal value near infinity is defined by omitting the divergent part of an integral as a function of the corresponding local parameter. \\
For $k=1,\dots,g$ we define:
\begin{align*}
{\bf 7.} \;\; \I_{r^k} [f(Q)] & := - \oint_{a_k} \l(Q) f_{\h}(Q) - \oint_{a_k} \lb(Q) f_{\ah}(Q) \\
{\bf 8.} \;\; \I_{u^k} [f(Q)] & := \oint_{b_k} \l(Q) f_{\h}(Q) + \oint_{b_k} \lb(Q) f_{\ah}(Q) \\
{\bf 9.} \;\; \I_{s^k} [f(Q)] & := \frac{1}{2\pi i} \oint_{b_k} f_\h(Q) \\
{\bf 10.} \;\; \I_{t^k} [f(Q)] & := - \frac{1}{2\pi i} \oint_{a_k} f_\h(Q) \;.
\end{align*}
Applying these operations to the sum of Schiffer and Bergman kernels, we shall obtain a set of primary differentials $\Phi\;,$ each of which gives a Darboux-Egoroff metric and a corresponding Frobenius structure. These differentials, listed below, decompose into a sum of holomorphic and antiholomorphic parts. The $a$-periods vanish for all primary differentials except for the differentials labeled by the index $s^k \;;$ the $b$-periods do not vanish only for the differentials having the index $t^k \;.$ This normalization and a given type of singularity characterize a primary differential completely due to the following lemma.
\begin{lemma}
\label{uniqueness}
If a single valued differential on a Riemann surface of the form $w=w_\h+w_{\ah}$ has zero $a$- and $b$-periods and its parts $w_\h$ and $w_{\ah}$ are everywhere analytic with respect to local parameters $z$ and $\bar{z} \,,$ respectively, then the differential $w$ is zero. 
\end{lemma}

{\it Proof.} Since the holomorphic and antiholomorphic parts of the differential must be regular and single valued on the surface, we can write $w$ in the form:
$ w = \sum_{k=1}^g \a_k \omega_k + \sum_{k=1}^g \b_k \overline{\omega_k} \; ,$
where $\{\omega_i\}$ are holomorphic normalized differentials. The vanishing of $a$-periods gives 
$\a_k=-\b_k$ and vanishing of $b$-periods implies that all $\a_k$ should be zero. 
${\Box}$ 

We list primary differentials together with their characteristic properties. 
A proof that the differentials have the given properties is essentially contained in the proof of Theorem \ref{thm_primary}.

Let us fix a point $P_0$ on $\surf$ such that  $\l(P_0)=0\;,$ and let all the basic cycles $\{a_k,b_k\}^g_{k=1}$ on the surface start at this point. This enables us to change the order of integration in expressions of the type $\oint_{b_k}\oint_{a_k}\l(P)\Omega(P,Q)$ (this can be checked by a local calculation of the integral near the point $P_0$) and compute $a$- and $b$-periods of the following primary differentials. 
\vspace{0.3cm}\\
For $i=0,\dots,m\;;\;\a=1,\dots,n_i\;:$
\begin{align*}
& {\bf 1.}  \;\;\Phi_{t^{i;\a}}(P) = \I_{t^{i;\a}} \left[ \Omega(P,Q) + B(\Pbar,Q) \right] & \sim (z_i^{-\a-1} + {\cal O}(1) ) dz_i + {\cal O}(1) d\bar{z}_i \;, \;\; P\sim\infty^i \;.\\
& {\bf 2.}  \;\; \Phi_{t^{\overline{i;\a}}}(P) = \overline{ \Phi_{t^{i;\a}}(P) } \;.          
\end{align*}
For $i=1,\dots,m\;:$ 
\begin{align*}
& {\bf 3.} \;\;\Phi_{v^i}(P) = \I_{v^i} \left[ \Omega(P,Q) + B(\Pbar,Q) \right] & {\sim} -d\l + {\cal O}(1) \left( dz_i + d\bar{z}_i \right) \;, \;\; P\sim\infty^i \;.\\
& {\bf 4.} \;\; \Phi_{v^{\ib}}(P) = \overline{\Phi_{v^i}(P)} \;. \\
&{\bf 5.}  \;\; \Phi_{w^i}(P) = \I_{w^i} \left[ \Omega(P,Q) + B(\Pbar,Q) \right] \;; & \underset{\infty^i}{\res} \; \Phi_{w^i} = 1 \;; \;\; \underset{\infty^0}{\res} \; \Phi_{w^i} = -1 \;. \qquad\qquad \\
& {\bf 6.}  \;\; \Phi_{w^{\ib}}(P) = \overline{\Phi_{w^i}(P)} \;.
\end{align*}
For $k=1,\dots,g\;:$ 
\begin{align*}
& {\bf 7.}  \;\; \Phi_{r^k}(P) = \I_{r^k} \left[ 2{\rm Re} \left\{  \Omega(P,Q) + B(\Pbar,Q) \right\} \right] \;; &&  \Phi_{r^k}(P^{b_k}) - \Phi_{r^k}(P) = 2\pi i d\l - 2\pi i d \lb \;.\\
& {\bf 8.} \;\; \Phi_{u^k}(P) = \I_{u^k} \left[ 2{\rm Re} \left\{ \Omega(P,Q) + B(\Pbar,Q) \right\} \right] \;; && \Phi_{u^k}(P^{a_k}) - \Phi_{u^k}(P) = 2\pi i d \l - 2 \pi i d \lb \;.  \\
& {\bf 9.} \;\; \Phi_{s^k}(P) = \I_{s^k} \left[ \Omega(P,Q) + B(\Pbar,Q) \right] \;; &&  \mbox{no singularities.} \\
& {\bf 10.} \;\; \Phi_{t^k}(P) = \I_{t^k} \left[ \Omega(P,Q) + B(\Pbar,Q) \right] \;; && \mbox{no singularities.}
\end{align*}
Here, as before, $\l=\l(P)$ and  $z_i=z_i(P)$ is the local parameter at $P \sim \infty^i$ such that $\l = z_i^{-n_i - 1}\;.$

Note that due to properties (\ref{S-B_periods}) of the Schiffer and Bergman kernels and the choice of the point $P_0$ (see the proof of Theorem \ref{thm_primary}), only the primary differentials of the last two types have non-zero $a$- and $b$-periods. Let us denote an arbitrary differential from the list by $\Phi_{\xi^A}\;;$ then the following holds:
\beqs
\oint_{a_\alpha}\Phi_{\xi^\iA}=\delta_{\xi^\iA,s^\alpha}\;;\qquad
\oint_{b_\alpha}\Phi_{\xi^\iA}=\delta_{\xi^\iA,t^\alpha}\;
\eeqs
($\delta$ is the Kronecker symbol). The number of primary differentials is $2L$ by virtue of the Riemann-Hurwitz formula (\ref{RH}).

Each of the primary differentials $\Phi$ defines a metric of the type (\ref{metric2}), (\ref{metric2'}) by the formula:
\beqn
{\bf ds}_\iPhi^{\bf 2} = \frac{1}{2} \sum_{i=1}^L \Phi_\h^2(P_i) (d\l_i)^2 + 
\frac{1}{2} \sum_{i=1}^L \Phi_{\ah}^2(P_i) (d\lb_i)^2 \;,
\label{Phimetric}
\eeqn
where $\Phi_\h$ and $\Phi_{\ah}$ are, respectively, the holomorphic and antiholomorphic parts of the differential $\Phi\;.$ The evaluation of differentials at a ramification point $P_i$ is done with respect to the standard local parameter $x_i = \sqrt{\l-\l_i}\;,$ i.e.  $\Phi_\h(P_i) = \left( \Phi_\h(P)/dx_i(P) \right)|_{P=P_i} \;.$ As is easy to see, metrics of the type (\ref{metric2'}) correspond to differentials $\Phi=\Phi_{u^k}$ and $\Phi=\Phi_{r^k}\;.$
\begin{proposition} Primary differentials satisfy the following relations:
\label{eOnPrim}
\beqn
{\bf e} \left( \Phi_\h (P_i) \right) = 0 \;, \qquad {\bf e} \left( \Phi_\ah (P_i) \right) = 0 \;,
\label{unit_prim}
\eeqn
for any ramification point $P_i\;.$ 
\end{proposition}
The proposition implies that the unit vector field {\bf e} (\ref{e2}) annihilates coefficients of the metric ${\bf ds}_\iPhi^{\bf 2}$ (\ref{Phimetric}).

{\it Proof.} 
Consider the covering $\cov^{\!\!\delta}$ obtained from $\cov$ by a $\delta$-shift of the points of every sheet, choosing $\delta \in \R\;;$ this shift maps the point $P$ of the surface to the point $P^\delta$ which belongs to the same sheet and for which $\l(P^\delta)=\l(P)+\delta \;.$ Denote by $\Omega^\delta$ and $B^\delta$ the corresponding kernels on $\cov^{\!\!\delta}\;.$ They are invariant with respect to biholomorphic mappings of the Riemann surface, i.e.
$\Omega^\delta(P^\delta,Q^\delta) = \Omega(P,Q) \;,$ and $ B^\delta(P^\delta,Q^\delta) = B(P,Q) \;.$
The local parameters near ramification points also do not change: $x_i(P) = x_i^\delta (P)   =\sqrt{\l(P) - \l_i}\;.$ Therefore for differentials $\Phi_{\omega^i} \;,\; \Phi_{\omega^{\ib}} \;,\; \Phi_{s^k}\;,$ and $\Phi_{t^k} \;,$ the statement of proposition follows immediately from this invariance. For them we have, for example,
\beqs
\Phi_{\omega^i \h}^\delta(P_j^\delta)=\Phi_{\omega^i \h}(P_j)\; ; \qquad 
\Phi_{\omega^i \ah}^\delta(P_j^\delta)=\Phi_{\omega^i \ah}(P_j)\; .
\eeqs
Differentiation of these equalities with respect to $\delta$ at $\delta = 0$ gives the action of the unit vector field ${\bf e}$ (\ref{e2}) on the differential in the left and zero in the right side.

Consider now the differential $\;\tilde{\Phi}(P) = - \oint_{a_k} \l(Q) \Omega(P,Q) - \oint_{a_k} \lb(Q) B(P,\bar{Q})\;,$  which is related to the differential $\Phi_{r^k}$ as follows:  $\Phi_{r^k}(P)  = 2 {\rm Re} \{ \tilde{\Phi}(P) \} \;.$ On the shifted covering $\cov^{\!\!\delta}$ we have
\begin{multline}
\tilde{\Phi}^\delta(P_i^\delta) =  - \oint_{a_k^\delta} \l(Q^\delta) \Omega^\delta(P^\delta_i,Q^\delta) - \oint_{a_k^\delta} \lb(Q^\delta) B^\delta(P^\delta_i,\bar{Q^\delta}) \\
= - \oint_{a_k} (\l(Q) + \delta) \Omega(P_i,Q) - \oint_{a_k} (\lb(Q) + \delta) B(P_i,\bar{Q}) \;.
\label{UnitPrim2}
\end{multline}
Differentiating both sides of this equality with respect to $\delta$ at $\delta = 0$ and using the property (\ref{S-B_periods}) of the Schiffer and Bergman kernels, we prove formulas (\ref{unit_prim}) for the differentials $\Phi_{r^k};$ the proof for $\Phi_{u^k}$ is analogous.

To prove (\ref{unit_prim}) for the remaining differentials consider the local parameter $z_i$ near infinity $\infty^i \;;$ under the $\delta$-shift it transforms as follows:
\beqs
z_i^{-\alpha}(P^\delta) \underset{\delta\sim 0}{=} (\l(P)+\delta)^\frac{\a}{n_i+1} = z_i^{-\a}(P) + \frac{\a}{n_i+1} ( z_i(P) )^{-\a +n_i + 1} \delta + {\cal O}(\delta^2) \;.
\eeqs
Therefore $\Phi_{t^{i;\a}}(P_j)$ on the covering $\cov^{\!\!\delta}$ is given by
\beqs
\Phi^\delta_{t^{i;\a}}(P_j^\delta) = \frac{1}{\a} \; \underset{\infty^i}{\res} \left( z_i^{-\a}(P) + \frac{\a}{n_i+1} (z_i(P))^{-\a + n_i + 1} \delta + {\cal O}(\delta^2) \right) \left( \Omega(P,P_j) +  B(P,\bar{P}_j) \right) \;.
\eeqs
Differentiating both sides with respect to $\delta$ at $\delta=0\;,$ we get 
\beqs
{\bf e} \left( \Phi_{t^{i;\a} \h} (P_j) \right) = \frac{1}{n_i + 1} \; \underset{\infty^i}{\res} \; (z_i(P))^{-\a + n_i + 1} \Omega(P,P_j) \;,
\eeqs
\beqs
{\bf e} \left( \Phi_{t^{i;\a} \ah} (P_j) \right) = \frac{1}{n_i + 1} \; \underset{\infty^i}{\res} \; (z_i(P))^{-\a + n_i + 1} B(P,\bar{P}_j)  \;.
\eeqs
The right sides are zero for non-negative powers of $z_i\;,$ i.e. for $\a=1, \dots, n_i + 1\;.$ This proves the statement of the proposition for differentials $\Phi_{t^{i;\a}}$ and $\Phi_{v^i}$ ( $\a = n_i + 1$ corresponds up to a constant to the case of differential $\Phi_{v^i}\;$).
$\Box$
\begin{remark}\rm This calculation also shows that differentials $\Phi_{t^{i;\a}}$ for  $i=0, \dots, m \;; \; \a=1, \dots, n_i$ and $\Phi_{v^i}$ for $i=1, \dots, m$  give the full set of primary differentials of the type $\oint_l \const \l^n (\Omega(P,Q) + B(P,Q))$ for $l$ being a small contour encircling one of the infinities. 

Note that we cannot consider $\Phi_{v^0}(P)$ as an independent differential due to the relation $\sum_{i=0}^m \Phi_{v^i}(P) = -(m+1)d\l(P)\;,$ where $d\l(P)=d\zeta$ is a differential on $\C P^1\;,$ the base of the covering. 
\end{remark}

Thus, we have constructed $2L$ differentials (see the Riemann-Hurwitz formula (\ref{RH})); each of them gives by formula (\ref{Phimetric}) a Darboux-Egoroff metric which satisfies {\bf F2} ($\nabla {\bf e} = 0$), and on which the Euler field acts according to (\ref{Euler2}) from {\bf F4}.

Our next goal is to find a set of flat coordinates for each of the metrics (\ref{Phimetric}).

\subsection{Flat coordinates}
Let us write the Christoffel symbols of the metric ${\bf ds}_\iPhi^{\bf 2}$ (\ref{Phimetric}) in terms of the corresponding  primary differential $\Phi\;.$ We shall use the following lemma which can be proven by a simple calculation using the definition of primary differentials and variational formulas (\ref{SB-variation}) for the Schiffer and Bergman kernels.
\begin{lemma} 
\label{primary_deriv}
The derivatives of primary differentials with respect to canonical coordinates are given by
\begin{align}
\frac{\d\Phi_{\xi^\iA}(P)}{\d\l_k} & = \frac{1}{2} \Phi_{\xi^\iA\h}(P_k) \left( \Omega(P,P_k) + B(\Pbar,P_k) \right) \\
\frac{\d\Phi_{\xi^\iA}(P)}{\d\lb_k} & = \frac{1}{2} \Phi_{\xi^\iA{\ah}}(P_k) \left( B(P,\Pbar_k) + \overline{\Omega(P,P_k)} \right) \;.
\end{align}
\end{lemma}
Then non-vanishing Christoffel symbols of the metric ${\bf ds}_\iPhi^{\bf 2}$ can be expressed as follows in terms of the primary differential $\Phi$ and rotation coefficients $\b_{ij}$ (\ref{SB-rotation}):
\begin{align}
\Gamma^j_{jk}  = \b_{jk} \frac{\Phi_\h(P_k)}{\Phi_\h(P_j)} &= -\Gamma_{kk}^i \;; \hsp \Gamma^j_{j\bar{k}}  = \b_{j\bar{k}} \frac{\Phi_{\ah}(P_k)}{\Phi_\h(P_j)} = -\Gamma_{\bar{k}\bar{k}}^j \;; \hsp
\Gamma^{\bar{j}}_{\bar{j}k} = \b_{\bar{j}k} \frac{\Phi_\h(P_k)}{\Phi_{\ah}(P_j)} = -\Gamma_{kk}^{\bar{j}} \;; 
\nonumber\\
\Gamma^{\bar{j}}_{\bar{j}\bar{k}} &= \b_{\bar{j}\bar{k}} \frac{\Phi_{\ah}(P_k)}{\Phi_{\ah}(P_j)}  = -\Gamma_{\bar{k}\bar{k}}^j \;; \qquad \Gamma_{jj}^j = - \sum_{l\neq j} \Gamma_{jl}^j \;.
\label{Christoffel}
\end{align}
Note that in the last formula, the index of summation $l$ runs through the set $\{1,\dots,L;\bar{1},\dots,\bar{L}\}\;.$

Flat coordinates can be found from the system of differential equation (\ref{flat}). Due to  formulas (\ref{Christoffel}), this system can be rewritten as follows:
\begin{align}
&\d_{\l_j} \d_{\l_k} t = \Gamma_{jk}^j \d_{\l_j} t + \Gamma_{jk}^k \d_{\l_k} t \;,\hsp j\neq k \in \{ 1, \dots, L, \bar{1}, \dots, \bar{L} \}
\label{flatCh1}\\
&{\bf e} (t) = {\rm const}\;.
\label{flatCh2}
\end{align}

Substituting expressions (\ref{Christoffel}) for Christoffel symbols into system (\ref{flatCh1}) and using Lemma \ref{primary_deriv}, one proves the next theorem by a straightforward computation. 
\begin{theorem} The following functions (and their linear combinations) satisfy system (\ref{flatCh1}):
\beqn
t_1 = \oint_{l_1} h_1 (\l(P)) \Phi_\h(P) \hsp {\mbox {and}} \hsp t_2 = \oint_{l_2} h_2(\lb(P)) \Phi_{\ah} (P) \;,
\label{solutions}
\eeqn
where $l_1\;,$ $l_2$ are two arbitrary contours on the surface $\surf$ which do not pass through ramification points and are such that their images $\l(l_1)$ and $\l(l_2)$ in $\zeta$-plane do not depend on $\{\l_k;\lb_k\}\;;$ arbitrary functions $h_1\;,$ $h_2$ are defined in some neighbourhoods of $l_1$ and $l_2\,,$ respectively, and are also independent of the coordinates $\{\l_k;\lb_k\}\;.$ The integration is regularized by omitting the divergent part where needed. 
\label{thmsolutions}
\end{theorem}

Among solutions (\ref{solutions}) we need to isolate those which satisfy equation (\ref{flatCh2}), the second part of the system identifying flat coordinates. The operations $\I_{\xi^\iA}$ applied to the differential $\Phi(P)$ give functions of the form (\ref{solutions}), and it turns out that flat coordinates can be obtained in this way. Namely, the following theorem holds.
\begin{theorem} 
\label{thm_flatcoord}
Let $P_0$ be a marked point on $\surf$ such that $\l(P_0)=0 \;.$ Let all the basic cycles $\{a_k,b_k\}_{k=1}^g$ start at the point $P_0\;.$
Then the following functions give a set of flat coordinates of the metric ${\bf ds}_\iPhi^{\bf 2}$ (\ref{Phimetric}).

For $i=0,\dots,m \;; \; \a=1,\dots,n_i \;:$ 
\beqs
t^{i;\a} := -(n_i+1) \I_{_{t^{i;1+n_i-\a}}}[\Phi] = \frac{n_i+1}{\a-n_i-1} \; \underset{\infty^i}{\res}\; z_i^{\a-n_i-1} \Phi_\h \;;
\eeqs
\beqs
t^{\overline{i;\a}} := -(n_i+1) \I_{_{t^{\overline{i;1+n_i-\a}}}}[\Phi] = \frac{n_i+1}{\a-n_i-1} \; \underset{\infty^i}{\tilde{\res}} \; \bar{z}_i^{\a-n_i-1} \Phi_\ah \;.
\eeqs
For $ i=1,\dots,m\;:$ 
\begin{alignat*}{5}
&v^i &&:= -\I_{w^i}[\Phi] &&= -\mathrm{v.p.} \int_{\infty^0}^{\infty^i} \Phi_\h \;; \hsp\hsp 
&v^{\bar{i}} &:= -\I_{w^{\bar{i}}}[\Phi] &&= -\mathrm{v.p.} \int_{\infty^0}^{\infty^i} \Phi_{\ah} \;;
\vspace{0.3cm}\\
&w^i &&:= -\I_{v^i}[\Phi] &&= - \; \underset{\infty^i}{\res} \; \l \Phi_\h \;; \hsp\hsp 
&w^{\bar{i}} &:= -\I_{v^{\bar{i}}}[\Phi] &&= -\; \underset{\infty^i}{\tilde{\res}} \; \lb \Phi_\ah \;.
\end{alignat*}
For $k=1,\dots,g\;:$
\begin{alignat*}{5}
&r^k &&:= \I_{s^k}[\Phi] &&= \frac{1}{2\pi i} \oint_{b_k} \Phi_\h \;; \hsp\hsp
&u^k &:= -\I_{t^k}[\Phi] &&= \frac{1}{2\pi i} \oint_{a_k} \Phi_\h \;;
\vspace{0.3cm}\\
%
&s^k &&:= \I_{r^k}[\Phi] &&= -\oint_{a_k} \left( \l \Phi_\h + \lb \Phi_\ah \right) \;; \hsp\hsp &t^k &:= - \I_{u^k}[\Phi] &&= -\oint_{b_k} \left( \l \Phi_\h + \lb \Phi_\ah \right) \;.
\end{alignat*}
As before, we use the notation $\tilde{\res}f := \overline{\res\bar{f}} \;.$

Let us denote the flat coordinates by
$\xi^A$ , i.e. we assume
\beqs
\xi^A\in\{t^{i;\a},t^{\overline{i;\a}}\;;\;v^i\;,\;v^{\ib}\;,\;w^{i}\;,\;w^{\ib}\;;\;r^k\;,\;u^k\;,\;s^k\;,\;t^k\}
\eeqs
for $\;i=0,\dots,m\;,\;\a=1,\dots,n_i\;;\;k=1,\dots,g $ (except $v^0,\,v^{\bar{0}}$ and $w^{0},\,\;w^{\bar{0}} \;,$ which do not exist).
\end{theorem}

{\it Proof.} Theorem \ref{thmsolutions} implies that these functions satisfy equations (\ref{flatCh1}).
The remaining equations (\ref{flatCh2}), ${\bf e} ( \xi^\iA ) = {\rm const}\;,$ can be proven by the same reasoning as in the proof of Proposition \ref{eOnPrim}. 
$\Box$ 

Note that the action of the unit vector field ${\bf e}$ (\ref{e2}) on a coordinate $\xi^\iA$ is non-zero if and only if the type of the coordinate coincides with the type of the primary differential which defines the metric. I.e. for the metric ${\bf ds}_\iPhi^{\bf 2}$ with $\Phi = \Phi_{\xi^{\iA_0}}$ the coordinate $\xi^{\iA_0}$ is naturally marked and we shall denote it by $\xi^1\;.$ One can prove that, for any choice of $\Phi\,,$ the corresponding coordinate $\xi^1$ is such that relations ${\bf e} (\xi^1) = -1$ and ${\bf e} (\xi^\iA) = 0$ hold for $\xi^\iA \neq \xi^1 \;.$ Therefore we have ${\bf e} = - \d_{\xi^1}$ (see also Proposition \ref{unityprop1} below).
\begin{remark}\rm By virtue of the Riemann-Hurwitz formula (see Section \ref{Hurwitz}), the number of functions listed in the theorem equals $2L\;,$ i.e. coincides with the number of canonical coordinates $ \{ \l_i; \lb_i \}. $  
\end{remark}

The next theorem gives an expression for the metric ${\bf ds}_\iPhi^{\bf 2}$ in coordinates $\{ \xi^\iA \}$ and by that shows again that functions $\{ \xi^\iA \left( \{ \l_k;\lb_k \} \right) \}$ are independent and play the role of flat coordinates of the metric. 
\begin{theorem}
\label{metricInFlat}
In coordinates $\{ \xi^\iA \}$ from Theorem \ref{thm_flatcoord} the metric ${\bf ds}_\iPhi^{\bf 2}$ (\ref{Phimetric}) is given by a constant matrix whose non-zero entries are the following:
\begin{align*}
& {\bf ds}_\iPhi^{\bf 2} \left( \d_{t^{i;\a}},\d_{t^{j;\b}} \right) = {\bf ds}_\iPhi^{\bf 2} \left( \d_{t^{\overline{i;\a}}},\d_{t^{\overline{j;\b}}} \right) = \frac{1}{n_i+1} \delta_{ij} \delta_{\a+\b, n_i+1} \;, \\
& {\bf ds}_\iPhi^{\bf 2} \left( \d_{v^i},\d_{\omega^j} \right) = {\bf ds}_\iPhi^{\bf 2} \left( \d_{v^{\ib}},\d_{\omega^{\jb}} \right) = \delta_{ij} \;, \\
& {\bf ds}_\iPhi^{\bf 2} \left( \d_{r^i},\d_{s^j} \right) = - \delta_{ij} \;, \\
& {\bf ds}_\iPhi^{\bf 2} \left( \d_{u^i},\d_{t^j} \right) = \delta_{ij} \;. 
\end{align*}
\end{theorem}
We shall prove this theorem later, after introducing a pairing of differentials (\ref{pairing}).

To further investigate properties of the flat coordinates let us choose one of the primary differentials $\Phi$ and build a multivalued differential on the surface $\surf$ as follows:
\beqn
\Psi(P)=\left(\vp\int_{\infty^0}^P\Phi_\h\right)d\l+\left(\vp\int_{\infty^0}^P\Phi_\ah\right)d\lb\;.
\label{mult}
\eeqn
This differential will play a role similar to the role of the differential $\p d\l$ in the construction of Dubrovin  (see formula (\ref{prepotential}) for prepotential). Note that $\Psi(P)$ decomposes into a sum of  holomorphic and antiholomorphic differentials: $\Psi=\Psi_{\h}+\Psi_{\ah} \;.$ 
\begin{theorem}
\label{PsiDeriv}
The derivatives of the multivalued differential $\Psi$ (\ref{mult}) with respect to flat coordinates $\{\xi^\iA\}$ are given by the corresponding primary differentials:
\beqs
\frac{\d\Psi}{\d\xi^\iA}=\Phi_{\xi^\iA}\;.
\eeqs
\end{theorem}

{\it Proof.}
Consider an expansion of the differential $\Psi$ in a neighbourhood of one of the infinities $\infty^i$ on the surface. We omit the singular part which does not depend on coordinates. As before, $z_i$ is a local coordinate in a neighbourhood of the  $i$-th infinity, $n_i$ is the corresponding ramification index. For $i\neq 0$ we have
\begin{multline}
\Psi(P) \underset{P\sim{\infty^i}}{=} {\mbox{singular part}} + \left( v^i(n_i+1) z_i^{-n_i-2} + \sum_{\alpha=1}^{n_i} t^{i;\alpha} z_i^{-\alpha-1} + w^iz_i^{-1} + {\cal O}(1) \right) dz_i \\
+ \left( v^{\ib}(n_i+1) \zb_i^{-n_i-2} + \sum_{\alpha=1}^{n_i} t^{\overline{i;\alpha}} \zb_i^{-\alpha-1} + w^{\ib} \zb_i^{-1} + {\cal O}(1) \right) d\zb_i \;.
\label{psi_expansion}
\end{multline}
We see that the expansion coefficients of the singular part are exactly the flat coordinates of the metric ${\bf ds_\Phi^2}\;.$ The coordinates $t^{0;\a}, \; \a = 1,\dots,n_0$ appear similarly in expansion at the infinity $\infty^0\;.$ The remaining coordinates $\xi^\iA$ correspond to other characteristics of the multivalued differential $\Psi\;.$ Namely, we have
\beqn
\oint_{a_k}\Psi=s^k\;,\qquad\qquad\oint_{b_k}\Psi=t^k\;;
\label{psi_periods}
\eeqn
\begin{align}
\Psi (P^{a_k})-\Psi(P) & = 2 \pi i u^k d\l - 2 \pi i u^k d\lb + \delta_{\scriptscriptstyle{\Phi,\Phi_{s^k}}} d\lb
+ \delta_{\scriptscriptstyle{\Phi,\Phi_{u^k}}} ( 2 \pi i d\l - 2 \pi i d\lb )
\label{psi_atwist} \;, 
\vspace{0.4cm}\\
\Psi(P^{b_k})-\Psi(P) & = 2 \pi i r^k d\l - 2 \pi i r^k d\lb + \delta_{\scriptscriptstyle{\Phi,\Phi_{t^k}}} d\lb
+ \delta_{\scriptscriptstyle{\Phi,\Phi_{r^k}}} ( 2 \pi i d\l - 2 \pi i d\lb ) \;,
\label{psi_btwist}
\end{align}
where $\Psi(P^{a_k})\;,$ $\Psi(P^{b_k})$ stand for the analytic continuation of $\Psi(P)$ along the corresponding cycles of the Riemann surface.

This parameterization of the differential $\Psi$ by the flat coordinates, together with Lemma \ref{uniqueness}, proves the theorem.
$\Box$

As a corollary we get the following lemma. 
\begin{lemma}
\label{lemmaJacobian}
The derivatives of canonical coordinates $\{\l_i\;;\lb_i\}$ with respect to flat coordinates $\{\xi^\iA\}$ of the metric ${\bf ds}_\iPhi^{\bf 2}$ are as follows
\beqs
\frac{\d\l_i}{\d\xi^\iA} = - \frac{\Phi_{\xi^\iA\h}(P_i)}{\Phi_\h (P_i)} \; , \qquad\qquad
\frac{\d\lb_i}{\d\xi^\iA} = - \frac{\Phi_{\xi^\iA\ah}(P_i)}{\Phi_\ah(P_i)} \; ,
\eeqs
where $\Phi(P)$ is the primary differential which defines the metric ${\bf ds}_\iPhi^{\bf 2} \;.$
\end{lemma} 

{\it Proof.} Theorem \ref{PsiDeriv} implies the following relations:
\beqn
\d_{\xi^\iA} \left\{ \left( \int^P_{\infty^0}  \Phi_\h \right) d\l \right\} = \Phi_{\xi^\iA\h} \;, \qquad \d_{\xi^\iA} \left\{ \left( \int^P_{\infty^0} \Phi_\ah \right)  d\lb \right\}= \Phi_{\xi^\iA\ah} \;.
\label{partsderiv}
\eeqn
(The divergent terms which we omit by taking the principal value of the integrals in a neighbourhood of $\infty^{\scriptscriptstyle{0}}$ do not depend on $\{\xi^\iA\}\;.$) 
We shall use the so-called thermodynamical identity 
\beqn
\d_\alpha(fdg)_{g=const}=-\d_\alpha (gdf)_{f=const}
\label{thermo}
\eeqn
for $f$ being a function of another function $g$ and some parameters $ \{ p_\a \}\,,$ i.e. $ f = f(g;p_1,\dots,p_n)\,,$ where $g$ can be expressed locally as a function of $f\;,$ i.e.   $g=g(f;p_1,\dots,p_n)\;;$ $\d_\alpha$ denotes the derivative with respect to one of the parameters $ p = \{ p_\a \} \;.$ Relation (\ref{thermo}) can be proven by differentiation of the identity
$f(g(f;p);p)\equiv f$ with respect to a parameter $p_\alpha \;,$ which gives $\d_\alpha gdf/dg+\d_\alpha f=0 \;.$
We use the thermodynamical identity (\ref{thermo}) for functions $f(P)=\int^P_{\infty^0}\Phi_\h$ and $g (P) = \l (P)$ to get
\beqs
 \d_{\xi^\iA} \left\{ \int^P_{\infty^0}\Phi_\h \right\} d\l = - \d_{\xi^\iA} \left\{ \l(P) \right\} \Phi_\h(P)\;,
\eeqs
and similarly, 
\beqs
 \d_{\xi^\iA} \left\{ \int^P_{\infty^0} \Phi_\ah \right\} d \lb = -  \d_{\xi^\iA} \left\{ \lb(P) \right\} \Phi_\ah (P) \;.
\eeqs
Evaluating these relations at the critical points $P=P_i \;,$ using that $\l^\prime(P_i)=0$ and equalities (\ref{partsderiv}), we prove the lemma.
$\Box$
\begin{proposition}
The unit vector field (\ref{e2}) is a tangent vector field in the direction of one of the flat coordinates. Namely, in flat coordinates of the metric ${\bf ds}_\iPhi^{\bf 2}$ (\ref{Phimetric}) corresponding to the primary differential $\Phi=\Phi_{\xi^{A_0}} \;,$ the unit vector of the Frobenius algebra is given by ${\bf e}=-\d_{\xi^{A_0}} \;.$ 
\label{unityprop1}
\end{proposition}

Let us denote the marked coordinate by $\xi^1$ so that ${\bf e}=-\d_{\xi^1}\;.$

{\it Proof.} This can be verified by a simple calculation using the chain rule $\d_{\xi^1} = \linebreak \sum_{i=1}^\iL \left( \frac{\d\l_i}{\d \xi^1}\d_{\l_i}+\frac{\d\lb_i}{\d \xi^1}\d_{\lb_i} \right)$
and expressions for ${\d\l_i}/{\d \xi^1}$ provided by Lemma \ref{lemmaJacobian}. 
$\Box$

\subsection{Prepotentials of new Frobenius structures}
A prepotential of the Frobenius structure which corresponds to a primary differential $\Phi$ is a function $F_\iPhi( \{ \xi^A \} )$ of flat coordinates of the metric ${\bf ds}_\iPhi^{\bf 2}$ such that its third derivatives are given by the tensor ${\bf c}$ from {\bf F3}:
\beqn
\frac{\d^3 F_\iPhi(\xi)}{\d{\xi^A}\d{\xi^B}\d{\xi^C}} = {\bf c}(\d_{\xi^A},\d_{\xi^B},\d_{\xi^C}) = {\bf ds}_\iPhi^{\bf 2} \left( \d_{\xi^A}\cdot\d_{\xi^B},\d_{\xi^C} \right) \;.
\label{prepot3der}
\eeqn
We shall construct a prepotential $F_\iPhi$ for each primary differential $\Phi\;.$ This will prove that {\bf F3} (symmetry of the tensor $(\nabla_{\xi^\iA} {\bf c})(\d_{\xi^\iB},\d_{\xi^\iC},\d_{\xi^\iD})$) holds in our construction. 
In order to write an expression for prepotential we define a new pairing of multivalued differentials as follows. 
 
Let $\omega^{(\a)}(P) \;, \; \a=1,2\dots$ be a differential on $\surf$ which can be decomposed into a sum of holomorphic ($\omega_{\h}^{(\a)}$) and antiholomorphic ($\omega_{\ah}^{(\a)}$) parts,
$\omega^{(\a)} = \omega_{\h}^{(\a)} + \omega_{\ah}^{(\a)} \; ,$ 
which are analytic outside infinities and have the following behaviour at $P\sim\infty^i\;$  (we write $\l$ for $\l(P)\;,$ and $z_i=z_i(P)$ for a local parameter $z_i^{-n_i-1}=\l$ at $P\sim\infty^i$ ):
\begin{align}
\begin{split}
\omega_{\h}^{(\a)}(P) & =  \sum _{n=-n^{(\a)}_1} ^\infty c_{n,i}^{(\a)} z_i ^ n dz_i + \frac{1}{n_i+1} d \left( \sum_{n>0} r_{n,i}^{(\a)} \l^n \log\l  \right)  \;, \\
\omega_{\ah}^{(\a)}(P) & = \sum_{n=-n^{(\a)}_2 }^\infty c_{\nb,i}^{(\a)} \zb_i^n d\zb_i + \frac{1}{n_i+1} d  \left( \sum_{n>0} r_{\nb,i}^{(\a)} \lb^n \log\lb \right) \;,
\label{coeff_def}
\end{split}
\end{align}
where $n_1^{(\a)}, n_2^{(\a)} \in \Z \;;$ and $c_{n,i}^{(\a)} \; , \; r_{n,i}^{(\a)} \; , \; c_{\bar{n},i}^{(\a)} \; , \; r_{\nb,i}^{(\a)}$ are some coefficients.
Denote also for $k = 1, \dots, g \;:$
\beqs
A_k^{(\a)} := \oint_{a_k}\omega^{(\a)}  \;, \qquad  B_k^{(\a)} := \oint_{b_k} \omega^{(\a)}  \;,
\eeqs
\begin{align}
\begin{split}
 dp_k^{(\a)}(\l(P)) \;  := \; \omega_{\h}^{(\a)}(P^{a_k}) - \omega_{\h}^{(\a)}(P) \;, \qquad &p_k^{(\a)} (\l)  = \sum_{s>0} p^{(\a)}_{sk} \l^s \;, \\
dp_{\kb}^{(\a)} (\bar{\l}(P))  := \; \omega_{\ah}^{(\a)}(P^{a_k}) - \omega_{\ah}^{(\a)}(P) \;, \qquad 
&p_{\kb}^{(\a)} (\bar{\l}) = \sum_{s>0} p^{(\a)}_{\bar{s}\kb}\bar{\l}^s \;,\\
dq_k^{(\a)}(\l(P)) := \; \omega_{\h}^{(\a)} (P^{b_k}) - \omega_{\h}^{(\a)}(P)  \;,
\qquad
&q_k^{(\a)} (\l) = \sum_{s>0} q^{(\a)}_{sk} \l^s \;,\\
dq_{\kb}^{(\a)}(\bar{\l}(P))  := \; \omega_{\ah}^{(\a)} (P^{b_k}) - \omega_{\ah}^{(\a)} (P)  \;,
\qquad
&q_{\kb}^{(\a)} (\bar{\l}) = \sum_{s>0} q^{(\a)}_{\bar{s}\kb} \bar{\l}^s \;.
\end{split}
\label{transformations}
\end{align}

Note that all primary differentials and the differential $\Psi(P)$ have singularity structures which are described by (\ref{coeff_def}) - (\ref{transformations}). For $\omega^{(\alpha)}$ being one of the primary differentials, the coefficients $c_{n,i}\,, \; r_{n,i}\,, $ $ \; c_{\nb,i}\,, \; r_{\nb,i}\,, $ $\; A_k\,, \; B_k\,, $ $ \; p_{sk}\,, \; q_{sk}\,, $ $\; p_{\bar{s} k}\,, \; q_{\bar{s} k}$  do not depend on coordinates on the Hurwitz space.

\begin{definition} 
For two differentials $\omega^{(\a)}\,,$ $\omega^{(\b)}$ having singularities of the type (\ref{coeff_def}), (\ref{transformations}), we define the {\bf pairing} $\F[\;,\;]$  as follows:
\begin{multline}
\label{pairing}
\F[ \omega^{(\a)}\;, \;\omega^{(\b)} ] = \sum_{i=0}^m \left( \sum_{n\geq 0} \frac{c^{(\a)}_{-n-2,i}}{n+1} c^{(\b)}_{n,i} + c_{-1,i}^{(\a)} \mathrm{v.p.} \int_{P_0}^{\infty^i} \omega_{\h}^{(\b)} - \mathrm{v.p.} \int_{P_0}^{\infty^i} \sum_{n>0} r_{n,i}^{(\a)} \l^n \omega_{\h}^{(\b)} \right. \\
 \hspace{2.5cm} + \left. \sum_{n\geq 0} \frac{c^{(\a)}_{-\overline{n-2},i}}{n+1} c^{(\b)}_{\nb,i} + c_{-\bar{1},i}^{(\a)} \mathrm{v.p.} \int_{P_0}^{\infty^i} \omega^{(\b)}_{\ah} - \mathrm{v.p.} \int_{P_0}^{\infty^i} \sum_{n>0} r_{\nb,i}^{(\a)} \lb^n \omega_{\ah}^{(\b)} \right) \\
  + \frac{1}{2\pi i} \sum_{k=1}^g \left( -\oint_{a_k} q_k^{(\a)}(\l) \omega_{\h}^{(\b)} + \oint_{a_k} q_{\kb}^{(\a)}(\bar{\l}) \omega_{\ah}^{(\b)} + \oint_{b_k}p_k^{(\a)}(\l) \omega_{\h}^{(\b)} \right. \\
\left. - \oint_{b_k} p_{\kb}^{(\a)}(\lb) \omega_{\ah}^{(\b)} + A_k^{(\a)} \oint_{b_k} \omega_{\h}^{(\b)} - B_k^{(\a)} \oint_{a_k} \omega_{\h}^{(\b)} \right) \;.
\end{multline}
As before, $P_0$ is the marked point on $\surf$ such that $\l(P_0)=0\;,$ and the cycles $\{a_k, b_k \}$ all pass through $P_0 \;.$
\end{definition}

From this definition one can see that if the first differential in the pairing is one of the primary differentials $\Phi_{\xi^\iA}$ then this pairing gives the corresponding operation $\I_{\xi^\iA}$ applied to the second differential:
\beqn
\label{pairing-action}
\F[\Phi_{\xi^\iA},\omega] = \I_{\xi^\iA}[\omega] \;.
\eeqn
\begin{theorem}
\label{thm_commutation}
The pairing (\ref{pairing}) is commutative for all primary differentials except for differentials $\Phi_{t^k}$ and $\Phi_{s^k}, \; k = 1,\dots, g$ which commute up to a constant:
\beqn
\label{pair_commut}
\F[\Phi_{s^k},\Phi_{t^k}] = \F[\Phi_{t^k},\Phi_{s^k}] - \frac{1}{2 \pi i} \;. 
\eeqn
\end{theorem}

{\it Proof.} Due to the relation (\ref{pairing-action}) we should compare the action of superpositions of operations $\I_{\xi^\iA}\I_{\xi^\iB}$ and $\I_{\xi^\iB}\I_{\xi^\iA}$ on the sum of Schiffer and Bergman kernels. This sum is only singular when the points $P$ and $Q$ coincide. Therefore among the operations $\I_{\omega^i}, \; \I_{\omega^{\ib}}, \; \I_{r^k}, \; \I_{u^k}, \; \I_{s^k}, \; \I_{t^k} $ those ones commute, being applied to $\Omega(P,Q) + B(P,Q)\;,$ which are given by integrals over non-intersecting contours on the surface. In the set of contours used in the definition of the operations $I_{\xi^\iA}\;,$ the only contours that intersect each other are the basis cycles $a_k$ and $b_k \;.$ A simple local calculation in a neighbourhood of the intersection point $P_0$ shows that the order of integration can be changed in the integral 
$\oint_{a_k}\oint_{b_k} \l(P) \Omega(P,Q)$
due to the assumption $\l(P_0)=0\;.$ Therefore the only non-commuting operations, among the mentioned above, are $\I_{s^k}$ and $\I_{t^k}\;.$ The difference in (\ref{pair_commut}) can be computed using formulas (\ref{W-periods}) for integrals of the bidifferential $W(P,Q)$ over $a$- and $b$-cycles.

By a similar reasoning one can see that operations of the type $\I_{t^{i;\a}} \;,$ $\I_{t^{{\ib};\a}} \;,$ $\I_{v^i} $ and $\I_{v^{\ib}}$ for $i=0,\dots,m \,, \; \a=1,\dots,n_i$ commute with the previous ones. They commute with each other due to the symmetry properties of the kernels. 
$\Box$ 

Now we are in a position to prove Theorem \ref{metricInFlat}, which gives the metric ${\bf ds}_\iPhi^{\bf 2}$ in flat coordinates.

{\it Proof of Theorem \ref{metricInFlat}.} For computation of the metric on vectors $\d_{\xi^\iA}$ we shall use the relation 
\beqn
\label{metr-pair}
{\bf ds}_\iPhi^{\bf 2} (\d_{\xi^\iA},\d_{\xi^\iB}) = {\bf e} \left( \F[\Phi_{\xi^\iA},\Phi_{\xi^\iB}] \right) \;,
\eeqn
which we prove first. 

Using Lemma \ref{lemmaJacobian},  we express the vectors $\d_{\xi^\iA}$ via canonical tangent vectors:
\beqn
\d_{\xi^\iA} = -\sum_{i=1}^L \left( \frac{\Phi_{\xi^\iA\h}(P_i)}{\Phi_{\h}(P_i)} \d_{\l_i} + \frac{\Phi_{\xi^\iA\ah}(P_i)}{\Phi_{\ah}(P_i)} \d_{\lb_i} \right) \;.
\label{flat_tangent}
\eeqn
Therefore for the metric (\ref{Phimetric}) we obtain:
\beqn
{\bf ds}_\iPhi^{\bf 2} (\d_{\xi^\iA},\d_{\xi^\iB}) = \frac{1}{2} \sum_{i=1}^L \left( \Phi_{\xi^\iA\h}(P_i) \Phi_{\xi^\iB\h}(P_i) + \Phi_{\xi^\iA\ah}(P_i) \Phi_{\xi^\iB\ah}(P_i) \right) \;.
\label{xiAxiBmetric}
\eeqn
For computation of the right-hand side of (\ref{metr-pair}) we note that, in the pairing of two primary differentials, only contribution of the second one depends on coordinates, therefore we have
\beqn
{\bf e} \left( \F[\Phi_{\xi^\iA},\Phi_{\xi^\iB}] \right) = \F[\Phi_{\xi^\iA},{\bf e} \left( \Phi_{\xi^\iB} \right)] \;. 
\label{temp3}
\eeqn
The action of the vector field ${\bf e}$ on primary differentials is provided by Lemma \ref{primary_deriv}. From (\ref{pairing-action}) we know that the pairing in the right side of (\ref{temp3}) is just the operation $\I_{\xi^\iA}$ applied to  ${\bf e} (\Phi_{\xi^\iB}) \;.$ Therefore in the right-hand side of (\ref{metr-pair}) we have
\begin{multline}
\frac{1}{2} \sum_{i=1}^L \left( \Phi_{\xi^\iB\h}(P_i) \I_{\xi^\iA} \left[ \Omega(P,P_i) + B(\bar{P},P_i) \right] + \Phi_{\xi^\iB\ah}(P_i) \I_{\xi^\iA} \left[ B(P,\bar{P_i}) + \overline{\Omega(P,P_i)} \right] \right) \\
= \frac{1}{2} \sum_{i=1}^L \left( \Phi_{\xi^\iA\h}(P_i) \Phi_{\xi^\iB\h}(P_i) + \Phi_{\xi^\iA\ah}(P_i) \Phi_{\xi^\iB\ah}(P_i) \right) \;.
\label{fun}
\end{multline}
Together with (\ref{xiAxiBmetric}), this proves (\ref{metr-pair}). 

Now let us compute ${\bf ds}_\iPhi^{\bf 2} \left( \d_{r^i}, \d_{\xi^\iA} \right)\;.$ According to (\ref{metr-pair}) we need to compute the action of the unit field ${\bf e}$ on the following quantity
\beqs
\F[\Phi_{r^i},\Phi_{\xi^\iA}] \equiv \I_{r^i}[\Phi_{\xi^\iA}] = - \oint_{a_i} \l(P) \Phi_{\xi^\iA\h}(P) - \oint_{a_i}  \lb(P) \Phi_{\xi^\iA\ah}(P) \;.
\eeqs
Let's again consider the biholomorphic map of the covering $\cov \to \cov^\delta$ performed by a simultaneous $\delta$-shift $(\delta \in \R)$ of the points on all sheets (see proof of Proposition \ref{eOnPrim}).
Since
\beqn
\label{epsilon}
\Phi^\delta_{\xi^\iA}(P^\delta) = \Phi_{\xi^\iA}(P)
\eeqn
we get 
\begin{multline*}
{\bf e} \left( \F[\Phi_{r^i},\Phi_{\xi^\iA}] \right) = \frac{d}{d \delta}|_{\delta = 0} \left( - \oint_{a_i} (\l(P) + \delta) \Phi_{\xi^\iA\h}(P) - \oint_{a_i} (\lb(P) + \delta) \Phi_{\xi^\iA\ah}(P) \right) \\
= - \oint_{a_i}\Phi_{\xi^\iA}(P) = - \delta_{\xi^\iA,s^i}\;.
\end{multline*}
Therefore ${\bf ds}_\iPhi^{\bf 2} \left( \d_{r^i}, \d_{\xi^\iA} \right) = - \delta_{\xi^\iA,s^i}\;.$ Analogously we prove that 
${\bf ds}_\iPhi^{\bf 2} \left( \d_{u^i}, \d_{\xi^\iA} \right) = \delta_{\xi^\iA,t^i} \;.$
To compute the remaining coefficients of the metric consider the operator
${\cal D}_{\bf e} = \frac{\d}{\d \l} + {\bf e} \;.$
It annihilates any primary differential:
\beqn
\label{De}
{\cal D}_{\bf e} \left( \Phi_{\xi^\iA}(P) \right) = 0
\eeqn
as can be proven by differentiation $d / d \delta |_{\delta = 0}$ of the equality (\ref{epsilon}). 
Therefore, applying the operator ${\cal D}_{\bf e} $ to the expansion of the multivalued differential $\Psi_{\xi^\iA}$ near the point $\infty^i\;,$ we obtain the following relation for the corresponding (see (\ref{coeff_def})) coefficients $c_{l,i}\;:$ 
\beqs
{\bf e} \left( c_{l,i}^{\scriptscriptstyle{\Phi_{\xi^\iA}}} \right) = \frac{l + 1}{n_i + 1} c_{l - n_i - 1,i}^{\scriptscriptstyle{\Phi_{\xi^\iA}}} \;.
\eeqs
Therefore we have
\beqs
{\bf ds}_\iPhi^{\bf 2} \left( \d_{t^{i;\a}}, \d_{\xi^\iA} \right) = {\bf e} \left( \I_{t^{i;\a}} [\Phi_{\xi^\iA}] \right) = {\bf e} \left( \frac{1}{\a} c_{\a-1,i}^{\scriptscriptstyle{\Phi_{\xi^\iA}}} \right) = \frac{1}{n_i + 1}\delta_{\xi^\iA,t^{i; 1 + n_i - \a}} \;,
\eeqs
and ${\bf ds}_\iPhi^{\bf 2} \left( \d_{v^i}, \d_{\xi^\iA} \right) = {\bf e} \left(  c_{n_i,i}^{\scriptscriptstyle{\Phi_{\xi^\iA}}} \right) = c^{\scriptscriptstyle{\Phi_{\xi^\iA}}}_{-1,i} = \delta_{\xi^\iA,\omega^i} \;.
$
Thus, we computed the entries of the matrix listed in the theorem and proved that they are the only non-zero ones. 
$\Box$ 

Formulas (\ref{flat_tangent}) and (\ref{xiAxiBmetric}) yield the following expression for the tensor ${\bf c}= {\bf ds}_\iPhi^{\bf 2}(\d_{\xi^\iA} \cdot \d_{\xi^\iB},\d_{\xi^\iC})$ (compare with expression (\ref{Dubrc_flat}) for the  tensor ${\bf c}$ of Dubrovin's construction):
\begin{multline}
{\bf c}(\d_{\xi^\iA},\d_{\xi^\iB},\d_{\xi^\iC}) = - \frac{1}{2}\sum_{i=1}^L \left( \frac{\Phi_{\xi^\iA\h}(P_i)\Phi_{\xi^\iB\h}(P_i)\Phi_{\xi^\iC\h}(P_i)}{\Phi_{\h}(P_i)} \right. \\
\left. + \frac{\Phi_{\xi^\iA\ah}(P_i)\Phi_{\xi^\iB\ah}(P_i)\Phi_{\xi^\iC\ah}(P_i)}{\Phi_{\ah}(P_i)}  \right)\;.
\label{c_flat}
\end{multline}

The next theorem gives a prepotential of the Frobenius manifold, a function of flat coordinates $\{ \xi^\iA \}\;,$ which, according to Theorem \ref{thm1}, solves the WDVV system. 
\begin{theorem} 
\label{main}
For each primary differential $\Phi$ consider the differential $\Psi(P)$ (\ref{mult}), multivalued on the surface $\surf\;.$ For the Frobenius structure defined on the manifold $\covM_{g;n_0,\dots,n_m}(\{ \l_i;\lb_i \})$ by the metric ${\bf ds}_\iPhi^{\bf 2}$ (\ref{Phimetric}), multiplication law (\ref{multiplicationR}), and Euler field (\ref{Eulerfield2}), the prepotential $F_\iPhi$ is given by the pairing (\ref{pairing}) of the differential $\Psi$ with itself:
\beqn
F_\iPhi=\frac{1}{2}\F [ \Psi\;,\; \Psi ]\;. 
\label{prepotential2}
\eeqn
The second order derivatives of the prepotential are given by
\beqn
\d_{\xi^A} \d_{\xi^B} F_\iPhi = \F [ \Phi_{\xi^A}\;,\;\Phi_{\xi^B} ] - \frac{1}{4 \pi i} \delta_{\xi^\iA,s^k}\delta_{\xi^\iB,t^k}   + \frac{1}{4 \pi i} \delta_{\xi^\iA,t^k}\delta_{\xi^\iB,s^k} \;,
\label{secondDer}
\eeqn
where $\delta$ is the Kronecker symbol.
\end{theorem}

{\it Proof.} To prove that the function $F_\iPhi$ is a prepotential we need to check that its third order derivatives coincide with the tensor ${\bf c}$ (\ref{c_flat}). We shall first prove that the second derivatives have the form (\ref{secondDer}) and then differentiate them with respect to a flat coordinate $\xi^\iC\;.$ 

The first differentiation of $F_\iPhi$ with respect to a flat coordinate gives:
\beqn
\d_{\xi^\iA} F_\iPhi = \frac{1}{2}\F[\Phi_{\xi^\iA},\Psi] + \frac{1}{2}\F[\Psi,\Phi_{\xi^\iA}]\;.
\label{der1}
\eeqn
The first term in the right side of (\ref{der1}) equals $\frac{1}{2}\I_{\xi^\iA}[\Psi]$ (see (\ref{pairing-action})). Consider the second term. 
From expansions (\ref{psi_expansion}) of the multivalued differential $\Psi$ and its integrals and transformations (\ref{psi_periods})-(\ref{psi_btwist}) over basis cycles we know that the coefficients for $\Psi$ which enter formula (\ref{pairing}) for the pairing are nothing but the flat coordinates of ${\bf ds}_\iPhi^{\bf 2} \;.$ Therefore, writing explicitly the singular part in expansions (\ref{psi_expansion}) and using also (\ref{psi_periods}) - (\ref{psi_btwist}), we have for the second term in (\ref{der1}):
\begin{multline}
\F[\Psi,\Phi_{\xi^\iA}]
 = \sum_{i=0}^m \left( v^i (1 - \delta_{i0}) \I_{v^i}[\Phi_{\xi^\iA}] + \sum_{\a=1}^{n_i} t^{i;\a} \I_{t^{i;\a}}[\Phi_{\xi^\iA}] + \omega^i \I_{\omega^i}[\Phi_{\xi^\iA}]  \right. \\
\left. + \delta_{\scriptscriptstyle{\Phi,\Phi_{\omega^i}}} \left( \frac{\I_{v^0}[\Phi_{\xi^\iA}]}{n_0+1} -  \frac{\I_{v^i}[\Phi_{\xi^\iA}]}{n_i+1} \right) + \delta_{\scriptscriptstyle{\Phi,\Phi_{\omega^i}}} \I_{\omega^i}[ \l \Phi_{\xi^\iA\h}] - \frac{1}{2} \delta_{\scriptscriptstyle{\Phi,\Phi_{v^i}}} \I_{v^i}[\l \Phi_{\xi^\iA\h}] \right. \\
\left. - \sum_{\a=1}^{n_i} \frac{\a(n_i+1)}{n_i+1+\a} \delta_{\scriptscriptstyle{\Phi,\Phi_{t^{i;\a}}}} \I_{t^{i;\a}}[\l \Phi_{\xi^\iA\h}] \right)\\
+ \sum_{i=0}^m \left( v^\ib (1 - \delta_{i0}) \I_{v^\ib}[\Phi_{\xi^\iA}] + \sum_{\a=1}^{n_i} t^{\overline{i;\a}} \I_{t^{\overline{i;\a}}}[\Phi_{\xi^\iA}] + \omega^\ib \I_{\omega^\ib}[\Phi_{\xi^\iA}] \right. \\
\left. + \delta_{\scriptscriptstyle{\Phi,\Phi_{\omega^\ib}}} \left( \frac{\I_{v^{\bar{0}}}[\Phi_{\xi^\iA}] }{n_0+1} - \frac{\I_{v^{\bar{i}}}[\Phi_{\xi^\iA}] }{n_i+1} \right) + \delta_{\scriptscriptstyle{\Phi,\Phi_{\omega^\ib}}} \I_{\omega^\ib} [ \lb \Phi_{\xi^\iA\ah} ] - \frac{1}{2} \delta_{\scriptscriptstyle{\Phi,\Phi_{v^\ib}}} \I_{v^\ib}[\lb \Phi_{\xi^\iA\ah}] \right. \\
\left. - \sum_{\a=1}^{n_i} \frac{\a (n_i+1)}{n_i+1+\a} \delta_{\scriptscriptstyle{\Phi,\Phi_{t^{i;\a}}}} \I_{t^{\overline{i;\a}}}[\lb \Phi_{\xi^\iA\ah}] \right) \\
+ \!\!\! \sum_{k=1} \! \left( \! r^k \I_{r^k}[\Phi_{\xi^\iA}] + u^k \I_{u^k}[\Phi_{\xi^\iA}] + s^k  \I_{s^k}[\Phi_{\xi^\iA}] + t^k \I_{t^k}[\Phi_{\xi^\iA}] + \frac{1}{2} \delta_{\scriptscriptstyle{\Phi,\Phi_{r^k}}} \! \I_{r^k} [\l\Phi_{\xi^\iA\h} \!\! + \!\! \lb\Phi_{\xi^\iA\ah}] \right. \\
\left. + \frac{1}{2} \delta_{\scriptscriptstyle{\Phi,\Phi_{u^k}}} \I_{u^k} [\l\Phi_{\xi^\iA\h} + \lb\Phi_{\xi^\iA\ah}]   + \frac{1}{2 \pi i} \delta_{\scriptscriptstyle{\Phi,\Phi_{s^k}}} \oint_{b_k} \lb \Phi_{\xi^\iA\ah} + \frac{1}{2 \pi i} \delta_{\scriptscriptstyle{\Phi,\Phi_{t^k}}} \oint_{a_k} \lb \Phi_{\xi^\iA\ah} \right)\;.
\label{multiline}
\end{multline}
Here the Kronecker symbol, for example, $\delta_{\scriptscriptstyle{\Phi,\Phi_{\omega^i}}}$ is equal to one if the primary differential $\Phi$ (which defines the metric ${\bf ds}_\iPhi^{\bf 2}$ and the differential $\Psi$) is $\Phi_{\omega^i}\;.$ 

Suppose the primary differential $\Phi_{\xi^\iA}$ is of the types $1,\; 3,\; 5,$ i.e. suppose $\xi^\iA \in \{ t^{i;\a},\; v^i,\; \omega^i \}\;.$ Then $\Phi_{\xi^\iA}(P) = \I_{\xi^\iA}[  \Omega(P,Q) + B(\bar{P},Q) ]\;.$ In this case the operation $\I_{\xi^\iA}$ commutes with all the others (see Theorem \ref{thm_commutation}). Therefore we can rewrite (\ref{multiline}) as an action of $\I_{\xi^\iA}$ on some differential which depends on $\l(Q)$ only (and does not depend on $\lb(Q)$): $\F[\Psi,\Phi_{\xi^\iA}] = \I_{\xi^\iA}[\tilde{\Psi}_\h(Q)]\;.$
Analogously, we find that for primary differentials of the types $2,\; 4,\; 6\;,$ when $\xi^\iA \in \{ t^{\overline{i;\a}},\; v^\ib,\; \omega^\ib \}\;,$ the right-hand side in (\ref{multiline}) is equal to the action of $\I_{\xi^\iA}$ on a differential depending only on $\lb(Q)\;,$ i.e. $\F[\Psi,\Phi_{\xi^\iA}] = \I_{\xi^\iA}[\tilde{\Psi}_\ah(Q)]\;.$
Examining the properties of the differential $\tilde{\Psi}_\h(Q) + \tilde{\Psi}_\ah(Q)$ such as singularities, behaviour under analytic continuation along cycles $\{ a_k,b_k \}$ and integrals over these cycles, we obtain with the help of Lemma \ref{uniqueness}: $\Psi(Q) = \tilde{\Psi}_\h(Q) + \tilde{\Psi}_\ah(Q)\;,$
and therefore $\Psi_\h(Q) = \tilde{\Psi}\h(Q) \;, \; \Psi_\ah(Q) = \tilde{\Psi}\ah(Q)\;.
$
Hence, for primary differentials of the types $1 - 6$ we have 
\beqn
\F[\Psi,\Phi_{\xi^\iA}] = \I_{\xi^\iA}[\Psi]\;.
\label{Psi_commut}
\eeqn
Similarly, for differentials $\Phi_{r^k}$ and $\Phi_{u^k}\;,$ we get
\begin{align*}
\F[\Psi,\Phi_{r^k}] 
&= - \oint_{a_k} \l(Q)\tilde{\Psi}_\h(Q) - \oint_{a_k} \lb(Q)\tilde{\Psi}_\ah(Q)   \;, \\
\F[\Psi,\Phi_{u^k}] &= - \oint_{b_k} \l(Q)\tilde{\Psi}_\h(Q) - \oint_{b_k} \lb(Q)\tilde{\Psi}_\ah(Q)   \;,
\end{align*}
which proves that (\ref{Psi_commut}) also holds for $\xi^\iA \in \{ r^k,\; u^k \} \;.$

Formula (\ref{Psi_commut}) changes for the primary differentials $\Phi_{s^k}$ and $\Phi_{t^k}\;:$ the additional terms appear due to non-commutativity of the corresponding operations (Theorem \ref{thm_commutation}):
\beqs
\F[\Psi,\Phi_{s^k}] = \I_{s^k}[\Psi] - \frac{t^k}{2 \pi i} \;; \qquad
\F[\Psi,\Phi_{t^k}] = \I_{t^k}[\Psi] + \frac{s^k}{2 \pi i} \;.
\eeqs
Coming back to the differentiation (\ref{der1}) of the function $F_\iPhi\;,$ we have 
\beqn
\d_{\xi^\iA} F_\iPhi = \F[\Phi_{\xi^\iA},\Psi] - \delta_{\xi^\iA,s^k} \frac{t^k}{4 \pi i} + \delta_{\xi^\iA,t^k} \frac{s^k}{4 \pi i}  \;.
\label{der2}
\eeqn
Note that the contribution of the primary differential $\Phi_{\xi^\iA}$ into the pairing $\F [ \Phi_{\xi^\iA},\Psi ]$ does not depend on coordinates. Therefore, by virtue of Theorem \ref{PsiDeriv}, the differentiation of (\ref{der2}) with respect to $\xi^\iB$ gives the expression (\ref{secondDer}) for second derivatives of the function $F_\iPhi\;.$ 

To find third derivatives of $F_\iPhi$ we differentiate (\ref{secondDer}) with respect to a flat coordinate $\xi^\iC\;:$
\beqn
\d_{\xi^\iC} \d_{\xi^B} \d_{\xi^A} F_\iPhi = \F [ \Phi_{\xi^A}\;,\;\d_{\xi^\iC} \Phi_{\xi^B} ] = \I_{\xi^\iA}[ \d_{\xi^\iC} \Phi_{\xi^B} ] \;.
\label{thirdDer}
\eeqn
Then we express the vector $\d_{\xi^\iC}$ via canonical tangent vectors $\{ \d_{\l_i} \}$ as in (\ref{flat_tangent}) and use formulas from Lemma \ref{primary_deriv} for derivatives of primary differentials. Analogously to the computation (\ref{fun}) we find that derivatives (\ref{thirdDer}) are given by the right-hand side of (\ref{c_flat}), i.e. equal to the $3$-tensor ${\bf c} (\d_{\xi^\iC},\d_{\xi^B},\d_{\xi^A})\;.$
$\Box$ 

Thus, by proving that the function $F_\iPhi$ given by (\ref{prepotential2}) is a prepotential (see Definition \ref{def_prepotential}) we completed the construction of Frobenius manifold corresponding to the primary differential $\Phi$ on the space $ \covM_{g;n_0,\dots,n_m}\;.$ Let us denote this manifold by $\covM^\iPhi = \covM^\iPhi_{g;n_0,\dots,n_m}\;.$

\subsection{Quasihomogeneity}

Now we shall show that the prepotential $F_\iPhi$ (\ref{prepotential2}) is a quasihomogeneous function of flat coordinates (see (\ref{quasihomogeneity})).
According to Theorem \ref{thm1}, the prepotential satisfies
\beqn
E (F_\iPhi) =  \nu_\iF F_\iPhi + \mbox{quadratic terms} \;.
\label{Euler_action}
\eeqn
In the next proposition we prove that the vector field $E$ has the form (\ref{Euler_flat}), i.e.
\beqn
E = \sum_{\iA}\nu_\iA \xi^\iA \d_{\xi^\iA} \;,
\label{Euler_flat_prop}
\eeqn
and compute the coefficients $\{ \nu_\iA \}\;.$
\begin{proposition}
\label{Prop_coeff_quasi}
In flat coordinates $\{ \xi^\iA \}$ of the metric ${\bf ds}_\iPhi^{\bf 2} \;,$ the Euler vector field (\ref{Eulerfield2}) has the form (\ref{Euler_flat_prop})
( and therefore is covariantly linear)
with coefficients $\{ \nu_\iA \}$ depending on the choice of a primary differential $\Phi$ as follows:
\begin{itemize}
\item if $\Phi = \Phi_{t^{i_o;\a}}$ or $\Phi = \Phi_{t^{\overline{i_o;\a}}}$ then
  	\begin{multline*}
	E = \sum_{i=0}^m \sum_{\a=1}^{n_i} \left( t^{i;\a} \d_{t^{i;\a}} + t^{\overline{i;\a}} 	\d_{t^{\overline{i;\a}}} \right) \left( 1+ \frac{\a}{n_{i_o}+1} - \frac{\a}{n_i+1} \right) \\
 	+ \sum_{i=1}^m \left( \frac{\a}{n_{i_o}+1} ( v^i \d_{v^i} + v^\ib \d_{v^\ib})  + (1+\frac{\a}{n_{i_o}+1}) ( \omega^i \d_{\omega^i}+ \omega^\ib \d_{\omega^\ib} ) 	\right) \\
	 + \sum_{k=1}^g \left( \frac{\a}{n_{i_o}+1} ( r^k \d_{r^k} +  u^k \d_{u^k} ) + 	(1+\frac{\a}{n_{i_o}+1}) ( s^k \d_{s^k} + t^k \d_{t^k} ) \right)
	\end{multline*}
\item if $\Phi = \Phi_{v^{i_o}}\;,$ $\Phi = \Phi_{v^{\ib_o}}\;,$  $\Phi = \Phi_{r^{k_o}}$ or $\Phi = \Phi_{u^{k_o}}$ then
  	\begin{multline*}
	E = \sum_{i=0}^m \sum_{\a=1}^{n_i}(2 - \frac{\a}{n_i+1}) ( t^{i;\a} \d_{t^{i;\a}} + t^{\overline{i;\a}} \d_{t^{\overline{i;\a}}} ) 
 + \sum_{i=1}^m \left( v^i \d_{v^i} + v^\ib \d_{v^\ib}
 + 2 ( \omega^i \d_{\omega^i} + \omega^\ib \d_{\omega^\ib} )t \right) \\
	 + \sum_{k=1}^g \left( r^k \d_{r^k} + u^k \d_{u^k} + 2 ( s^k \d_{s^k} +  t^k \d_{t^k} ) \right)
	\end{multline*}
\item if $\Phi = \Phi_{\omega^{i_o}}\;,$ $\Phi = \Phi_{\omega^{\ib_o}}\;,$ $\Phi = \Phi_{s^{k_o}}$ or $\Phi = \Phi_{t^{k_o}}$ then
  	\beqs
	E = \sum_{i=0}^m \sum_{\a=1}^{n_i}(1 - \frac{\a}{n_i+1}) ( t^{i;\a} \d_{t^{i;\a}} + t^{\overline{i;\a}} \d_{t^{\overline{i;\a}}} ) + \sum_{i=1}^m ( \omega^i \d_{\omega^i} +  \omega^\ib \d_{\omega^\ib} )
	 + \sum_{k=1}^g  ( s^k \d_{s^k} + t^k \d_{t^k} ) \;.
	\eeqs
\end{itemize}
\end{proposition}

{\it Proof.} Let us compute the action of the Euler vector field on a flat coordinate $\xi^\iA\;.$ Consider again the biholomorphic map $\cov \to \cov^{\!\!\epsilon} $ defined by the transformation $P\mapsto P^\epsilon$ on $\surf$ such that $\l(P^\epsilon) = \l(P)(1+\epsilon)\;, \; \epsilon \in \R\;,$ performed on every sheet of the covering $\cov \;.$ Since the kernels $\Omega$ and $B$ are invariant under this map, the primary differentials transform as follows:
\begin{alignat*}{2}
& \; {\rm for} \;\;  \Phi = \Phi_{t^{i;\a}}\;\; {\rm or} \;\; \Phi = \Phi_{t^{\overline{i;\a}}}\;: & \Phi^\epsilon (P^\epsilon) = (1 + \epsilon)^{\frac{\a}{n_i + 1}} \Phi(P) \\
&  \; {\rm{for} } \;\; \Phi = \Phi_{v^i},\;\; \Phi = \Phi_{v^\ib},\;\; \Phi = \Phi_{r^k}\;\; {\rm or} \;\; \Phi = \Phi_{u^k}: & \Phi^\epsilon (P^\epsilon) = (1 + \epsilon) \Phi(P)\\
&  \; {\rm for } \;\; \Phi = \Phi_{\omega^i}, \;\; \Phi = \Phi_{\omega^\ib}, \;\; \Phi = \Phi_{s^k}\;\; {\rm or} \;\; \Phi = \Phi_{t^k}: & \Phi^\epsilon (P^\epsilon) = \Phi(P) \;,
\end{alignat*}
where $\Phi^\epsilon$ is the corresponding differential on the covering $\surf_\l^\epsilon\;.$

Let us choose, for example, the primary differential $\Phi_{t^{{i_o};\a}}\;.$ Flat coordinates of the metric ${\bf ds}_{\scriptscriptstyle{\Phi_{t^{{i_o};\a}}}}^{\bf 2}$ are functions of $\{ \l_j \}$ and $\{ \lb_j \}$ only. If we consider corresponding functions on $\surf^\epsilon$ and differentiate them with respect to $\epsilon$ at  $\epsilon = 0\;,$ we get the action of the vector field $E$ (\ref{Eulerfield2}) on the flat coordinates:
\begin{multline*}
E(t^{i;\a})  = \frac{d}{d\epsilon}|_{\epsilon = 0}  \frac{n_i+1}{\a-n_i-1} \; \underset{\infty^i}{\res} \; ( \l(P^\epsilon) )^{\frac{n_i+1-\a}{n_i+1}} \Phi^\epsilon_{t^{{i_o};\a}\h}(P^\epsilon) 
\\= \frac{d}{d\epsilon}|_{\epsilon = 0} (1+\epsilon)^{\frac{n_i +1-\a}{n_i +1} + \frac{\a}{n_{i_o} +1}}t^{i;\a} =( 1 - \frac{\a}{n_i+1} + \frac{\a}{n_{i_o}+1} ) t^{i;\a}  .
\end{multline*}
Therefore the vector field $E$ depends on the coordinate $t^{i;\a}$ as $E = ( 1 - \frac{\a}{n_i+1} + \frac{\a}{n_{i_o}+1} ) t^{i;\a} \d_{t^{i;\a}} + \dots .$ Similarly we compute the dependence on the other flat coordinates. 
$\Box$ 

The action (\ref{Euler_action}) of the Euler field (\ref{Euler_flat_prop}) on the prepotential $F_\iPhi$ is equivalent to the condition of quasihomogeneity for $F_\iPhi\;,$ i.e. $F_\iPhi (\kappa^{\nu_1}\xi^1,\dots, \kappa^{\nu_{2\iL}}\xi^{2\iL})= \kappa^{\nu_\iF} F_\iPhi(\xi^1,\dots,\xi^{2\iL}) + quadratic\; terms$
with the coefficients of quasihomogeneity $\{ \nu_\iA \}$ computed in Proposition \ref{Prop_coeff_quasi}. As for the coefficient $\nu_\iF\;,$ the proof of Theorem \ref{thm1} implies that $\nu_\iF = 3-\nu\;,$ where the charge $\nu$ of a Frobenius manifold was computed in Proposition \ref{propEulermetric2}. Thus, we have
\begin{alignat*}{3}
& \; {\rm for} \;\;  \Phi = \Phi_{t^{i;\a}}\;\; {\rm or} \;\; \Phi = \Phi_{t^{\overline{i;\a}}}\;: & \nu = 1 - \frac{2\a}{n_i+1} &\qquad \nu_\iF = \frac{2\a}{n_i+1} +2 \\
&  \; {\rm for } \;\; \Phi = \Phi_{v^i},\;\; \Phi = \Phi_{v^\ib},\;\; \Phi = \Phi_{r^k}\;\; {\rm or} \;\; \Phi = \Phi_{u^k}: &  \nu = -1 & \qquad \nu_\iF = 4 \\
&  \; {\rm for } \;\; \Phi = \Phi_{\omega^i}, \;\; \Phi = \Phi_{\omega^\ib}, \;\; \Phi = \Phi_{s^k}\;\; {\rm or} \;\; \Phi = \Phi_{t^k}: & \nu = 1 & \qquad \nu_\iF = 2 \;.
\end{alignat*}
\begin{remark}\rm The described construction also holds for the differential $\Phi$ being a linear combination of the primary differentials which correspond to the same charge $\nu\;.$ In other words, the differential $\Phi$ which defines a Frobenius structure can be one of the following:
\begin{align*}
{\bf 1.} \;\; & \Phi = c_{i;\a} \Phi_{t^{i;\a}} + c_{\overline{i;\a}} \Phi_{t^{\overline{i;\a}}} \qquad \mbox{for some pair } (i;\a): \;\; i \in \{0,\dots,m\} \;, \;\;  \a \in \{1,\dots,n_i-1\} \;, \\
{\bf 2.} \;\; & \Phi = \sum_{i=1}^m \left( \kappa_i\Phi_{v^i} + \kappa_\ib\Phi_{v^\ib} \right) + \sum_{k=1}^g \left( \sigma_k\Phi_{r^k} + \rho_k\Phi_{u^k} \right) \;, \\
{\bf 3.} \;\; & \Phi = \sum_{i=1}^m \left( \kappa_i\Phi_{\omega^i} + \kappa_\ib\Phi_{\omega^\ib} \right) + \sum_{k=1}^g \left( \sigma_k\Phi_{s^k} + \rho_k\Phi_{t^k} \right) \;,
\end{align*}
where the coefficients do not depend on a point of the Hurwitz space. The unit vector fields for the structures defined by these combinations, respectively, are given by:
\begin{align*}
{\bf 1.} \;\; & {\bf e} = - c_{i;\a} \d_{t^{i;\a}} - c_{\overline{i;\a}} \d_{t^{\overline{i;\a}}} \qquad \mbox{for some pair } (i;\a): \;\; i=0,\dots,m \;, \;\;  \a=1,\dots,n_i-1 \\
{\bf 2.} \;\; & {\bf e} = - \sum_{i=1}^m \left( \kappa_i \d_{v^i} + \kappa_\ib \d_{v^\ib} \right) - \sum_{k=1}^g \left( \sigma_k\d_{r^k} + \rho_k\d_{u^k} \right) \\
{\bf 3.} \;\; & {\bf e} = - \sum_{i=1}^m \left( \kappa_i\d_{\omega^i} + \kappa_\ib \d_{\omega^\ib} \right) - \sum_{k=1}^g \left( \sigma_k\d_{s^k} + \rho_k\d_{t^k} \right)\;. 
\end{align*}
In each case, by a linear change of variables, the field ${\bf e}$ can be made equal to $\d_{\xi^1}$ for some new variable $\xi^1\;.$ This change of variables does not affect the quasihomogeneity of the prepotential since the flat coordinates which enter each of the three combinations have equal coefficients of quasihomogeneity (see Proposition \ref{Prop_coeff_quasi}).
\label{remark_comb}
\end{remark}

\section{$G$-function of Hurwitz Frobenius manifolds}
\label{G-function}

The $G$-function is a solution to the Getzler system of linear differential equations, which was derived in \cite{Getzler} (see also \cite{DubZhang}). The system is defined on an arbitrary semisimple Frobenius manifold $M\;.$ 

It was proven in \cite{DubZhang} that the Getzler system has unique, up to an additive constant, solution $G$ which satisfies the quasihomogeneity condition 
\beqs
E(G) = - \frac{1}{4} \sum _{\iA = 1}^\dim \left( 1-\nu_\iA - \frac{\nu}{2} \right)^2 +\frac{ \nu\dim}{48}  \;,
\eeqs
with a constant in the left side: $\nu$ is the charge, $\dim$ is the dimension of the Frobenius manifold; $\{\nu_\iA \}$ are the quasihomogeneity coefficients (\ref{quasihomogeneity}). In \cite{DubZhang} the following formula (which proves the conjecture of A. Givental \cite{Givental}) for this quasihomogeneous solution was derived:
\beqn
G = \log \frac{\tau_\iI}{J^{\scriptscriptstyle{1/24}}}  \;,
\label{G-funct}
\eeqn
where $J$ is the Jacobian of transformation from canonical to the flat coordinates,
$J = \det \left( \frac{\d t^\alpha}{\d \l_i} \right);$
and $\tau_\iI$ is the isomonodromic tau-function of the Frobenius manifold defined by 
\beqn
\frac{\d \log \tau_\iI}{\d \l_i} = H_i := \frac{1}{2} \sum_{j \neq i, j = 1} ^\dim \b_{ij}^2 (\l_i - \l_j) 
\;, \qquad i = 1, \dots, \dim \;.
\label{tauiso}
\eeqn
The function $G$ (\ref{G-funct}) for the Frobenius manifold $\covM_{1;1}^{\phi_s}$ was computed in \cite{DubZhang}. In \cite{KokKorB, KokKorG} expression (\ref{G-funct}) was computed for Dubrovin's Frobenius structures on Hurwitz spaces in arbitrary genus. Theorem \ref{KokKorthm} below summarizes the main results of papers \cite{KokKorB} and \cite{KokKorG}.  

Denote by $S$ the following term in asymptotics of the bidifferential $W(P,Q)$ (\ref{W-def}) near the diagonal $ P \sim Q \;:$
\beqs
W(P,Q) \underset{Q \sim P}{=} \left( \frac{1}{(x(P) - x(Q))^2} + S(x(P)) + o(1) \right) dx(P) dx(Q)
\eeqs
( $6S(x(P))$ is called the Bergman projective connection \cite{Fay92}).
By $S_i$ we denote the value of $S$ at the ramification point $P_i$ taken with respect to the local parameter $x_i(P) = \sqrt{ \l - \l_i }\;:$
\beqn
S_i = S(x_i)|_{x_i = 0} \;.
\label{Si}
\eeqn
Since the singular part of the bidifferential $W$ in a neighbourhood of the point $P_i$ does not depend on coordinates $\{ \l_j \}\;,$ the Rauch variational formulas (\ref{W-variation}) imply
\beqs
\frac{\d S_i}{\d \l_j} = \frac{1}{2} W^2(P_i,P_j) \;.
\eeqs
The symmetry of this expression provides compatibility for the following system of differential equations which defines the Bergman tau-function $\tau_\iW\;:$
\beqs
\frac{\d \log \tau_\iW }{\d \l_i} = - \frac{1}{2} S_i \;, \qquad  i=1, \dots, \dim \; .
\eeqs
\begin{theorem}
\label{KokKorthm}
The isomonodromic tau-function $\tau_\iI $ (\ref{tauiso}) for a holomorphic Frobenius structure $\covM^\phi$ is related to the Bergman tau-function $\tau_\iW$ as follows (\cite{KokKorG}):
\beqn
\tau_\iI = (\tau_\iW)^{-\frac{1}{2}} \;,
\label{tau-tau}
\eeqn
where $\tau_\iW$ is given by the following expression independent of the points $P$ and $Q$ (\cite{KokKorB}):
\begin{equation}
\label{tauW}
\tau_\iW = {\cal Q}^{2/3} \prod_{k,l=1 \; k<l}^{\iL+m+1} \left[ E(D_k,D_l) \right]^{d_k d_l/6}
\end{equation}
and 
\itemize
\item ${\cal Q}$ is given by
\beqs
{\cal Q} = \left[ d\l(P) \right]^{\frac{g-1}{2}} {\cal C}(P) \prod_{k=1}^{\iL+m+1} \left[ E(P,D_k) \right]^\frac{(1-g) d_k}{2}
\eeqs
where ${\cal C}(P)$ is the following multivalued $g(1-g)/2$-differential on $\surf$
\beqs
{\cal C}(P) = \frac{1}{ \det_{1 \leq \a, \b \leq g} \| \omega_\b ^{(\a - 1)} (P) \|} \sum^g_{\a_1, \dots, \a_g = 1} \frac{\d^g \theta(K^P)}{\d z_{\a_1} \dots \d z_{\a_g}} \omega_{\a_1}(P) \dots \omega_{\a_g}(P) 
\eeqs
\item $\sum_{k=1}^{\iL+m+1} d_k D_k$ is the divisor 
$(d\l)$ of the differential $d\l(P)\;,$ i.e. $D_l =P_l, \; d_l=1$ for $l = 1, \dots, L$ and $D_{\iL+i+1} = \infty^i, \; d_{\iL+i+1} = -(n_i+1), \; i = 0,\dots, m \;.$ As before, we evaluate a differential at the points of the divisor $(d\l)$ with respect to the standard local parameters: $x_j = \sqrt {\l-\l_j}$ for $j=1,\dots,L$ and $x_{\iL+1+i} = \l^{-1/(n_i+1)}$ for $i=0,\dots,m$
\item $\theta(z|\B), \; z \in \C^g$ is the theta-function; $E(P,Q)$ is the prime form;  $E(D_k,P)$  stands for $E(Q,P) \sqrt{d x_k (Q)}|_{Q=D_k}$
\item $K^P$ is the vector of Riemann constants; the fundamental domain $\widehat{\surf}$ is chosen so that the Abel map of the divisor $(d\l)$ is given by ${\cal A}((d\l))= -2K^\iP \;.$ 
\end{theorem}

\subsection{G-function for manifolds $\covM^\phi$}

Theorem \ref{KokKorthm} gives the numerator of expression (\ref{G-funct}) for the $G$-function of holomorphic Frobenius structures $\covM^\phi$ on Hurwitz spaces described in Section \ref{DubFrob}. For the denominator we have (see \cite{DubZhang}, \cite{KokKorG})
\beqs
J = \frac{1}{2^{\iL/{\scriptscriptstyle 2}}} \prod_{i=1}^L \phi(P_i) \;,
\eeqs
where $\phi$ is the primary differential from the list of Theorem \ref{thm_primary} which corresponds to the Frobenius structure $\covM^\phi\;.$ 

Summarizing above formulas, we get the following expression for the $G$-function of the Frobenius manifold $\covM^\phi\;:$
\beqn
G = - \frac{1}{2} \log \tau_\iW - \frac{1}{24} \log \prod_{i=1}^L \phi(P_i) + {\rm const} \;,
\label{G-funct1}
\eeqn
$\tau_\iW$ is given by (\ref{tauW}).

\subsection{G-function for ``real doubles" $\covM^\iPhi$}

For the Frobenius structures with canonical coordinates $\{ \l_1, \dots, \l_L; \lb_1, \dots, \lb_L \}\;,$ corresponding to the primary differentials $\Phi$ from Section \ref{NewFrob}, the Jacobian of transformation between canonical and flat coordinates is given by
\beqn
J = \det \left( \frac{\d \xi^\iA}{\d \l_i} \Big{|} \frac{\d \xi^\iA}{\d \lb_i} \right) =  \frac{1}{2^\iL} \prod_{i=1}^L \Phi_\h(P_i) \Phi_\ah(P_i) \;.
\label{Jacobian2}
\eeqn
The definition (\ref{tauiso}) of the isomonodromic tau-function in this case becomes:
\begin{align}
\begin{split}
\frac{\d \log \tau_\iI}{\d \l_i} = H_i := \frac{1}{2} \sum _{j \neq i, j=1}^L \beta_{ij}^2 (\l_i - \l_j) + \frac{1}{2} \sum _{j=1}^L \beta_{i\jb}^2 (\l_i - \lb_j)  \\
\frac{\d \log \tau_\iI}{\d \lb_i} = H_{\overline{i}} := \frac{1}{2} \sum _{j=1}^L \beta_{\bar{i}j}^2 (\lb_i - \l_j) + \frac{1}{2} \sum _{j \neq i, j=1}^L \beta_{\bar{i}\jb}^2 (\lb_i - \lb_j) \;.
\label{tauiso2}
\end{split}
\end{align}
Analogously to relation (\ref{tau-tau}) one can prove (see \cite{KokKorG} and Proposition \ref{tau-tau-prop} below) that the function $\tau_\iI$ is $-1/2$ power of the function $\tau_\iOmega \;,$ which is defined by the Schiffer kernel $\Omega(P,Q)$ (\ref{Omegadef}) as follows. The asymptotics of the kernel $\Omega(P,Q)$ near the diagonal is 
\beqs
\Omega(P,Q) \underset{Q \sim P}{=} \left( \frac{1}{(x(P) - x(Q))^2} + S_\iOmega(x(P)) + o(1) \right) dx(P) dx(Q) \;.
\eeqs
Denote by $\Omega_i$ the evaluation of the term $S_\iOmega(x)$ at the ramification point $P_i$ with respect to the local parameter $x_i = \sqrt{\l-\l_i} \;:$
\beqs
\Omega_i = \left( S_\iOmega(x_i)\right)|_{x_i=0} = S_i+\Sigma_i \;,
\eeqs
where $S_i$ is the same as in (\ref{Si}) and $\Sigma_i$ is given by $\Sigma_i = - \pi \sum_{k,l = 1}^g ({\rm Im}\B)^{-1}_{kl} \omega_k(P_i) \omega_l(P_i)  
\;.$ 
The differentiation formulas (\ref{SB-variation}) for the kernels $\Omega$ and $B$ imply
\begin{alignat}{2}
\begin{split}
\frac{\d \Omega_i}{\d \l_j} = \frac{1}{2} \Omega^2(P_i,P_j) = 2 \beta_{ij}^2 \;, \qquad & 
\frac{\d \Omega_i}{\d \lb_j} = \frac{1}{2} B^2(\bar{P}_j,P_i) = 2 \beta_{i\jb}^2 \;, \\
\frac{\d \overline{\Omega}_i}{\d \l_j} = \frac{1}{2} B^2(\bar{P}_i,P_j) = 2 \beta_{\bar{i}j}^2 \;, \qquad & 
\frac{\d \overline{\Omega}_i}{\d \lb_j} = \frac{1}{2} \overline{\Omega^2(P_i,P_j)} = 2 \beta_{\bar{i}\jb}^2 \;,
\label{m_deriv}
\end{split}
\end{alignat}
which allows the following definition of the tau-function $\tau_\iOmega\;:$
\beqn
\frac{\d \log \tau_\iOmega}{\d \l_i} = - \frac{1}{2} \Omega_i \;, \qquad 
\frac{\d \log \tau_\iOmega}{\d \lb_i} = - \frac{1}{2} \overline{\Omega}_i \;.
\label{tauOmega}
\eeqn
From Rauch variational formulas (\ref{Rauch}) 
we find 
\beqs
\frac{\d \log \det ({\rm Im} \B)}{\d \l_i} = - \frac{1}{2} \Sigma_i \;, \qquad
\frac{\d \log \det ({\rm Im} \B)}{\d \lb_i} = - \frac{1}{2} \overline{\Sigma}_i \;,
\eeqs
and therefore
\beqn
\tau_\iOmega = {\rm const} \; | \tau_\iW |^2 \det ({\rm Im} \B) \;. 
\label{tauO-tauW}
\eeqn
\begin{remark}\rm This tau-function coincides with an appropriately regularized ratio of determinant of Laplacian on $\surf$ and the volume of $\surf$ in the singular metric $|d\l|^2$ (see \cite{D'HokerPhong, KokKorB, Sonoda}).
\end{remark}

Now we are able to compute the function $\tau_\iI$ (\ref{tauiso2}) by proving the following proposition. 
\begin{proposition}
\label{tau-tau-prop}
The isomonodromic tau-function $\tau_\iI$ for a Frobenius structure with canonical coordinates $\{ \l_1,\dots, \l_L; \lb_1,\dots,\lb_L \}$ on the Hurwitz space is related to the function $\tau_\iOmega$ (\ref{tauOmega}) by 
\beqn
\tau_\iI = (\tau_\iOmega)^{-1/2} \;.
\label{tau-tau2}
\eeqn
\end{proposition}

{\it Proof.} Using the relation (\ref{m_deriv}) between derivatives of $\Omega_i$ and rotation coefficients $\b_{ij}\;,$ we write for the Hamiltonians $H_i$ (\ref{tauiso2}):
\beqn
H_i \! = \! \frac{1}{4} \l_i \left( \sum_{j \neq i, j=1}^L \d_{\l_j} \Omega_i + \! \sum_{j=1}^L \d_{\lb_j}\Omega_i \right) - \frac{1}{4} \!\! \sum_{j \neq i, j=1}^L \l_j \d_{\l_j} \Omega_i
 - \frac{1}{4} \! \sum_{j=1}^L \lb_j \d_{\lb_j}\Omega_i \;.
\label{tempo}
\eeqn
For the quantities $\Omega_i$ one can prove the relations
\beqn
\sum_{j=1}^L \left( \frac{\d}{\d \l_j} + \frac{\d}{\d \lb_j} \right)  \Omega_i = 0 \;,
\qquad \sum_{j=1}^L \left( \l_j \frac{\d}{\d \l_j} + \lb_j \frac{\d}{\d \lb_j} \right)  \Omega_i = -\Omega_i \;.
\label{tempo1}
\eeqn
To prove (\ref{tempo1}) we use the invariance of the Schiffer kernel $\Omega(P,Q)$ under two biholomorphic maps of the Riemann surface $\surf \mapsto \surf^\delta$ and $\surf \mapsto \surf^\epsilon$ given by transformations $\l \to \l+\delta$ and $\l \to \l(1+\epsilon)$ performed simultaneously on all sheets of the covering $\cov$ (see proofs of Propositions \ref{metric2-prop} and \ref{propEulermetric}).

Substitution of (\ref{tempo1}) into (\ref{tempo}) yields 
\beqs
H_i = - \frac{1}{4} \sum_{j=1}^L \left( \l_j \d_{\l_j}\Omega_i + \lb_j \d_{\lb_j}\Omega_i \right) = \frac{1}{4}\Omega_i \;.
\eeqs
Similarly, we get for $H_{\bar{i}}$ the relation:
$H_{\bar{i}} = \frac{1}{4}\overline{\Omega}_i \;.$
$\Box$ 

Formulas (\ref{Jacobian2}), (\ref{tauO-tauW}) and (\ref{tau-tau2}) give the expression for the function $G$ (\ref{G-funct}), i.e. we have proven the following theorem.
\begin{theorem}
The $G$-function of the Frobenius manifold $\covM^{\Phi}$ is given by
\beqn
G = - \frac{1}{2} \log \left\{  \; | \tau_W |^2 \det ({\rm Im} \B)  \right\} - \frac{1}{24} \log \left\{ \prod_{i=1}^L \Phi_\h(P_i) \Phi_\ah(P_i) \right\} + {\rm const} \;,
\label{G-funct2}
\eeqn
where the Bergman tau-function $\tau_\iW$ is given by (\ref{tauW}).
\end{theorem}
\section{Examples in genus one}
\label{Examples}
Since the described construction in the case of genus zero does not lead to new structures, the simplest examples we can compute are the Frobenius structures in genus one. The simplest Hurwitz space in genus one is $M_{1;1}\;.$ We shall compute the prepotentials of Frobenius manifolds $\covM_{1;1}^{\phi_s}$ and $\covM_{1;1}^{\Phi_s}\;,$  $\covM_{1;1}^{\Phi_t}\;,$ $\covM_{1;1}^{\Phi_s+\sigma\Phi_t}$ (for a nonzero constant $\sigma \in \C$) given by formulas (\ref{prepotential}) and (\ref{prepotential2}), respectively, and the corresponding $G$-functions (\ref{G-funct1}) and (\ref{G-funct2}). 

The Riemann surface of genus one can be represented as a quotient $\surf = \C / \{ 2\omega, 2\omega^\prime \} \;,$ where $ \omega, \omega^\prime \in \C \;.$  
The space $M_{1;1}$ consists of the genus one two-fold coverings of $\C P^1$ with simple branch points, one of them being at infinity. These coverings can be defined by the function 
\beqn
\l(\z) = \wp(\z) + c \;,
\label{P-torus}
\eeqn
where $\wp$ is the Weierstrass  elliptic function $\wp:\surf \to \C P^1$ and  $c$ is a constant with respect to $\z \;.$

We denote by $\l_1,\l_2,\l_3$ the finite branch points of the coverings (\ref{P-torus}) and consider them as local coordinates on the space $\covM_{1;1}\;.$ 

\subsection{Holomorphic Frobenius structure $\covM_{1;1}^{\phi_s}$ }

The primary differential $\phi_s$ is the holomorphic normalized differential (see (\ref{W-periods})):
\beqn
\phi(\z)=\phi_s(\z)=\frac{1}{2\pi i}\oint_bW(\z,\tilde{\z}).
\eeqn
It can be expressed as follows via $\l$ and $\z\;:$
\beqn
\phi(\l(\z))=\frac{1}{4\omega}\frac{d\l}{\sqrt{(\l-\l_1)(\l-\l_2)(\l-\l_3)}} \qquad\qquad 
\phi(\z)=\frac{d\z}{2\omega}\;.
\eeqn
The expansion of multivalued differential $pd\l=\left(\int_0^\z\phi\right)d\l$  at infinity with respect to the local parameter $z=\l^{-1/2}$ is given by
\beqs
pd\l=\frac{1}{2\omega}\left(\frac{2}{z^2}+c+\mathcal{O}(z)\right)dz \;.
\eeqs
The Darboux-Egoroff metric (\ref{phimetric}) corresponding to our choice of primary differential $\phi$ has in canonical coordinates $\{\l_i\}$ the form
\beqn
{\bf ds}_{\phi_s}^{\bf 2} = \frac{1}{8\omega^2} \left\{ \frac{(d\l_1)^2}{(\l_1-\l_2)(\l_1-\l_3)} + \frac{(d\l_2)^2}{(\l_2-\l_1)(\l_2-\l_3)} + \frac{(d\l_3)^2}{(\l_3-\l_1)(\l_3-\l_2)} \right\} \;.
\label{metric_can}
\eeqn
The set of flat coordinates of this metric is
\begin{align}
\begin{split}
&t_1 := s = - \oint_a \l\phi_s = - \frac{1}{2\omega} \int_x^{x+2\omega} (\wp(\z)+c) d\z = 
-\frac{\pi i}{4\omega^2}\gamma - c \\
& t_2 := t^{\scriptscriptstyle{0;1}} = \; \underset{\z = 0} {\res} \; \frac{1}{\sqrt{\l}} pd \l = \frac{1}{\omega} \\
& t_3 := r = \frac{1}{2\pi i} \oint_b \phi = \frac{1}{2\pi i} \frac{\omega^\prime}{\omega} \;,
\label{flatcoord-ex}
\end{split}
\end{align}
where we denote by $\gamma$ the following function of period $\mu=2\pi i t_3$  of the torus $\surf\;:$
\begin{equation}
\gamma(\mu) = \frac{1}{3\pi i} \frac{\theta_1^{\prime\prime\prime}(0;\mu)}{\theta_1^\prime(0;\mu)} \;.
\label{gamma}
\end{equation}
This function satisfies the Chazy equation (see for example \cite{Dubrovin}):
\beqn
\gamma^{\prime\prime\prime} = 6 \g \g^{\prime\prime} - 9 \gamma^{\prime\; 2} \;.
\eeqn
The metric (\ref{metric_can}) in coordinates (\ref{flatcoord-ex}) is constant and has the form:
\beqs
{\bf ds}_{\phi_s}^{\bf 2} = \frac{1}{2} (dt_2)^2 - 2 dt_1 dt_3 \;.
\eeqs
The prepotential (\ref{prepotential}) (it was computed in \cite{Bertola, Dubrovin}) of the Frobenius structure $\covM_{1;1}^{\phi_s}$
is given by
\beqs
F_{\phi_s} = - \frac{1}{4}t_1 t_2^2 + \frac{1}{2}t_1^2t_3 - \frac{\pi i}{32} \; t_2^4 \; \gamma(2\pi i t_3)\;. 
\eeqs
This function is quasihomogeneous, i.e. the following relation
\begin{equation}
F_{\phi_s} (\kappa^{\nu_1} t_1, \kappa^{\nu_2} t_2, \kappa^{\nu_3} t_3 ) = \kappa^{\nu_\iF} F_{\phi_s}(t_1, t_2, t_3)
\label{quasihom}
\end{equation}
holds for any $\kappa \neq 0$ and the quasihomogeneity factors 
\begin{equation}
\nu_1 = 1 \;, \qquad \nu_2 = \frac{1}{2}\;,\qquad \nu_3 = 0 \qquad {\mbox{and}}\qquad \nu_\iF = 2 \;.
\label{quasihomcoeff}
\end{equation}
The Euler vector field $E=\sum_{i=1}^3\l_i\d_{\l_i}$ in flat coordinates has the form:
\beqs
E = \sum_{k=1}^3 \nu_\alpha t_\alpha \d_{t_\alpha} = t_1\d_{t_1} + \frac{1}{2} t_2 \d_{t_2} \;;
\eeqs
and the quasihomogeneity (\ref{quasihom}), (\ref{quasihomcoeff}) can be written as 
$E(F_{\phi_s}(t_1, t_2, t_3)) = 2 F_{\phi_s}(t_1, t_2, t_3)\;.\;\;$

The corresponding $G$-function was computed in \cite{DubZhang} :
\beqs
G = - \log \left\{ \eta(2 \pi i t_3) (t_2)^\frac{1}{8} \right\} + {\rm const} \;,
\eeqs
where $\eta(\mu)$ is the Dedekind eta-function: $\eta(\mu) = (\theta^\prime_1(0))^{1/3}\;.$ (See \cite{KokKorB} for the function $\tau_\iW$ in genus one.)

\subsection{ ``Real doubles" in genus one }

We consider the same coverings $(\surf,\l)$ with $\surf = \C / \{ 2\omega, 2\omega^\prime \} \;, $ and the function $\l$ given by $(\ref{P-torus})\;.$ The coverings have simple branch points $\l_1,\l_2,\l_3$ and $\infty\;.$ The set of such coverings is considered now as a space with local coordinates  $\{\l_1,\l_2,\l_3;\lb_1,\bar{\l}_2,\bar{\l}_3\}\;.$ 

\subsubsection{ The manifold $\covM_{1;1}^{\Phi_s}$ }

The primary differential $\Phi=\Phi_s$  has the form ($\mu=\omega^\prime/\omega$ is the period of the torus $\surf$):
\beqn
\Phi(\z) = \Phi_s(\z) = \frac{\bar{\mu}}{\bar{\mu} - \mu} \frac{d\z}{2\omega} + \frac{\mu}{\mu-\bar{\mu}} \frac{\overline{d\z}}{2\bar{\omega}} \;.
\label{Phi_s}
\eeqn
The corresponding Darboux-Egoroff metric (\ref{Phimetric}) is given by 
\begin{multline}
  {\bf ds}_{\Phi_s}^{\bf 2} \!  = \! {\rm Re} \left\{\frac{1}{4\omega^2}\left(\frac{\bar{\mu}}{\bar{\mu} - \mu}\right)^2\!\!\!   \left(\frac{(d\l_1)^2}{(\l_1 - \l_2)(\l_1 - \l_3)} +\frac{(d\l_2)^2}{(\l_2-\l_1)(\l_2-\l_3)}  \right.\right. \\
 +  \left.\left. \frac{(d\l_3)^2}{(\l_3-\l_1)(\l_3-\l_2)}\right)\right\} \;.
\label{metric1s_can}
\end{multline}
The flat coordinates of this metric are 
\begin{alignat}{2}
\label{flatPhi_s}
t_1 & := s = {\rm Re} \left\{ \frac{\bar{\mu}}{\mu-\bar{\mu}} \int_x^{x+2\omega} \!\!\!\! (\wp(\z)+c) \frac{d\z}{\omega} \right\}
    &\qquad t_4 & := t = {\rm Re} \left\{ \frac{\bar{\mu}}{\mu-\bar{\mu}} \int_x^{x+2\omega^\prime} \!\!\!\!\! (\wp(\z)+c) \frac{d\z}{\omega} \right\}
\nonumber\\
t_2 & := t^{\scriptscriptstyle{0;1}} = \frac{\bar{\mu}}{\bar{\mu}-\mu}\frac{1}{\omega} 
    & \qquad t_5 & := t^{\overline{\scriptscriptstyle{0;1}}} = \bar{t}_2 \\
t_3 & := r = \frac{1}{2\pi i} \frac{\mu\bar{\mu}}{\bar{\mu}-\mu}
    &\qquad t_6 & := u = \frac{1}{2\pi i} \frac{\bar{\mu}}{\bar{\mu}-\mu} \;. \nonumber
\end{alignat}
Note that $ \mu = t_3/t_6\;,$ $ \bar{\mu} = 2 \pi i t_3/(2 \pi i t_6 - 1)$ 
and for the solution (\ref{gamma}) to the Chazy equation we have  $\overline{\g(\mu)} = - \g(-\bar{\mu})\;.$ 

The metric (\ref{metric1s_can}) in the flat coordinates has the form
\beqs
{\bf ds}_{\Phi_s}^{\bf 2}=\frac{1}{2}(dt_2)^2+\frac{1}{2}(dt_5)^2-2dt_1dt_3+2dt_4dt_6\;.
\eeqs

The corresponding prepotential (\ref{prepotential2}) is 
\begin{align}
 \begin{split}
   F_{\Phi_s}  & = -\frac{1}{4} t_1 t_2^2 - \frac{1}{4} t_1 t_5^2 + \frac{1}{2} t_1^2 t_3 - \frac{1}{2} t_1 t_4 (2 t_6 - \frac{1}{2 \pi i} ) \\
   & + t_3^{-1} \left( \frac{1}{4} t_2^2 t_4 (t_6 - \frac{1}{2 \pi i} ) + \frac{1}{4}t_4 t_5^2 t_6 + \frac{1}{2} t_4^2 t_6 ( t_6 - \frac{1}{2 \pi i}) + \frac{1}{16} t_2^2 t_5^2  \right)  \\
   & + 
   \frac{1}{32} t_2^4 \left( - \frac{1}{4 \pi i} t_6^{-2} \; \gamma \left( \frac{t_3}{t_6} \right) + t_3^{-1} -    \frac{1}{2\pi i} t_3^{-1}t_6^{-1} \right)  \\
   & +  
   \frac{1}{32} t_5^4 \left( -\frac{\pi i} {(2 \pi i t_6 - 1)^2} \; \gamma \left( \frac{2 \pi i t_3}{1 - 2 \pi i t_6}    \right) + t_3^{-1} + t_3^{-1} (2 \pi i t_6 - 1)^{-1} \right).   
 \label{Fs}
 \end{split}
\end{align}
Note that the coordinates $ t_1, \; t_3, \; t_4$ are real, $t_2$ and $t_5$ are complex conjugates of each other and $t_6$ has a constant imaginary part, $\bar{t}_6 = t_6 - 1 / 2 \pi i\;.$ In these coordinates, the prepotential $F_{\Phi_s} $ is a real-valued function. However, $F_{\Phi_s}$ also satisfies the WDVV system when considered as a function of six complex coordinates; in that case, $F_{\Phi_s}$ is not real.

This function is quasihomogeneous: the relation $F_{\Phi_s} (\kappa^{\nu_1} t_1,\dots, \kappa^{\nu_6} t_6)= \kappa^{\nu_\iF} F_{\Phi_s} (t_1, \dots, t_6)$ holds for any $\kappa \neq 0$ and the quasihomogeneity factors 
\begin{align}
\begin{split}
\nu_1=1\;, &\qquad \nu_2=\frac{1}{2}\;,\qquad \nu_3=0\;,  \\
\nu_4 = 1\;, &\qquad \nu_5 = \frac{1}{2}\;, \qquad \nu_6 = 0\;, \qquad \nu_\iF=2\;.
\label{quasihomcoeff1}
\end{split}
\end{align}
The Euler vector field $E=\sum_{i=1}^3(\l_i\d_{\l_i} + \lb_i\d_{\lb_i})$  has the following form in the flat coordinates:
\beqs
E=\sum_{\alpha=1}^6 \nu_\alpha t_\alpha \d_{t_\alpha} = t_1\d_{t_1} + \frac{1}{2} t_2 \d_{t_2} + t_4\d_{t_4} + \frac{1}{2} t_5 \d_{t_5}\;,
\eeqs
and the quasihomogeneity of $F_{\Phi_s}$ can be written as 
$E(F_{\Phi_s} (t_1, \dots, t_6))= 2 F_{\Phi_s} (t_1,\dots, t_6)\;.$

The corresponding $G$-function (\ref{G-funct2}) (real-valued as a function of coordinates (\ref{flatPhi_s})) is given by 
\beqs
G =  - \log \left\{ \eta \left( \frac{t_3}{t_6} \right) \eta \left( \frac{2 \pi i t_3}{1- 2 \pi i t_6} \right) \left( t_2 t_5 \right)^\frac{1}{8} \left( \frac{ 2 \pi i t_3}{t_6(2 \pi i t_6 - 1 )} \right)^\frac{1}{2}  \right\} + {\rm const} \;.
\eeqs
Here we use the relation  $\overline{\eta(\mu)} = \eta(-\bar{\mu})$  for the Dedekind $\eta$-function. 

\subsubsection { The manifold $\covM_{1;1}^{\Phi_t}$ }

The primary differential $\Phi=\Phi_t$  has the form ($\mu=\omega^\prime/\omega$ is the period of torus):
\beqn
\Phi(\z) = \Phi_t(\z) = \frac{1}{\mu - \bar{\mu}} \frac{d\z}{2\omega} - \frac{1}{\mu-\bar{\mu}} \frac{\overline{d\z}}{2\bar{\omega}} \;.
\label{Phi_t}
\eeqn
The corresponding Darboux-Egoroff metric (\ref{Phimetric}) is given by 
\begin{multline}
  {\bf ds}_{\Phi_t}^{\bf 2} \!  = \! {\rm Re} \left\{\frac{1}{4\omega^2}\left(\frac{1}{\bar{\mu} - \mu}\right)^2\!\!\!   \left(\frac{(d\l_1)^2}{(\l_1 - \l_2)(\l_1 -                                 \l_3)} + \frac{(d\l_2)^2}{(\l_2-\l_1)(\l_2-\l_3)}  \right.\right. \\
 +  \left.\left. \frac{(d\l_3)^2}{(\l_3-\l_1)(\l_3-\l_2)}\right)\right\} \;.
\label{metric1t_can}
\end{multline}
The flat coordinates of this metric are 
\begin{alignat}{2}
\label{flatPhi_t}
t_1 & := t = {\rm Re} \left\{ \frac{1}{\bar{\mu} - \mu} \int_x^{x+2\omega^\prime} \!\!\!\!\! (\wp(\z)+c) \frac{d\z}{\omega} \right\}
 &\qquad t_4 & := s = {\rm Re} \left\{ \frac{1}{\bar{\mu} - \mu} \int_x^{x+2\omega} \!\!\!\! (\wp(\z)+c) \frac{d\z}{\omega} \right\} \nonumber\\
t_2 & := t^{\scriptscriptstyle{0;1}} = \frac{1}{\mu - \bar{\mu}} \frac{1}{\omega} 
    & \qquad t_5 & := t^{\overline{\scriptscriptstyle{0;1}}} = \bar{t}_2 \\
t_3 & := r = \frac{1}{2\pi i} \frac{\mu}{\mu - \bar{\mu}}
    &\qquad t_6 & := u = \frac{1}{2\pi i} \frac{1}{\mu - \bar{\mu}} \nonumber
\end{alignat}
In terms of these coordinates, the period of the torus and its conjugate can be expressed as:  $ \mu = t_3/t_6$ and $ \bar{\mu} = (2 \pi i t_3-1)/2 \pi i t_6 \;.$

The metric (\ref{metric1t_can}) in flat coordinates has the form:
\beqs
{\bf ds}_{\Phi_t}^{\bf 2} = \frac{1}{2}(dt_2)^2 + \frac{1}{2}(dt_5)^2 + 2 dt_1 dt_6 - 2 dt_3 dt_4\;.
\eeqs

The corresponding prepotential (\ref{prepotential2}) is given by
\begin{align}
 \begin{split}
   F_{\Phi_t}  & = -\frac{1}{4} t_1 t_2^2 - \frac{1}{4} t_1 t_5^2 + \frac{1}{2} t_1 t_4 (2 t_3 - \frac{1}{2 \pi i} ) - \frac{1}{2} t_1^2 t_6  - \frac{1}{2} t_3 (t_3 - \frac{1}{2 \pi i}) \frac{t_4^2}{t_6} - \frac{1}{16} \frac{t_2^2 t_5^2}{t_6}   \\
   & - 
   \frac{t_2^4}{32 t_6} - \frac{1}{128 \pi i} \frac{t_2^4}{t_6^{2}} \; \gamma \left( \frac{t_3}{t_6} \right) + \frac{t_3 t_4 t_5^2}{4 t_6}  \\
   & - 
   \frac{t_5^4}{32 t_6} - \frac{1}{128 \pi i} \frac{t_5^4}{t_6^{2}} \; \gamma \left( \frac{1 - 2 \pi i t_3}{2 \pi i t_6} \right) + \frac{(t_3 - \frac{1}{2 \pi i}) t_4 t_2^2}{4 t_6}   \;.
 \label{Ft}
 \end{split}
\end{align}
This function is also real if the coordinates are of the form (\ref{flatPhi_t}): in this case $ t_1, \; t_4, \; t_6$ are real, $t_2 = \bar{t}_5\;,$ and $t_3$ has a constant imaginary part, namely, we have $\bar{t}_3 = t_3 - \frac{1}{2 \pi i}\;.$ Last two lines in (\ref{Ft}) are complex conjugates of each other since for the function $\gamma$ (\ref{gamma}) we have $\overline{\gamma(\mu)} = -\gamma(-\bar{\mu})\;.$

The function $F_{\Phi_t}$ (\ref{Ft}) is quasihomogeneous. The quasihomogeneity factors $\{ \nu_i \}$ and $\nu_\iF$ are the same as for the above example (the function $F_{\Phi_s}$), they are given by (\ref{quasihomcoeff1}). 

The $G$-function for $\covM_{1;1}^{\Phi_t}$ (it is also real-valued as a function of coordinates (\ref{flatPhi_t})) is given by
\beqs
G = - \log \left\{ \eta \left( \frac{t_3}{t_6} \right) \eta \left( \frac{1 - 2 \pi i t_3}{ 2 \pi i t_6} \right) \left( t_2 t_5 \right)^\frac{1}{8} t_6^{-\frac{1}{2}}  \right\} + {\rm const} \;,
\eeqs
where, again, $\eta$ is the Dedekind eta-function.

\subsubsection { The manifold $\covM_{1;1}^{\Phi_s+\sigma\Phi_t}$ }

According to Remark \ref{remark_comb} in the end of Section \ref{NewFrob}, there exists a Frobenius structure built from a linear combination of two primary differentials $\Phi_s$ and $\Phi_t\;.$ Here, we compute a prepotential which corresponds to the differential $\Phi = \Phi_s+\sigma\Phi_t$ for $\sigma$ being a non-zero parameter. 

We start with the differential
\beqs
\Phi(\z) = \Phi_s(\z) + \sigma \Phi_t(\z) = \frac{\bar{\mu}-\sigma}{\bar{\mu}-\mu} \frac{d\z}{2\omega} + \frac{\sigma-\mu}{\bar{\mu}-\mu} \frac{\overline{d\z}}{2\bar{\omega}} \;.
\eeqs

The corresponding Darboux-Egoroff metric (\ref{Phimetric}) is given by
\begin{multline}
  {\bf ds}_{\Phi}^{\bf 2} \!  = \! \frac{1}{8\omega^2}\left(\frac{\bar{\mu}-\sigma}{\bar{\mu}-\mu} \right)^2\!\!\!   \left(\frac{(d\l_1)^2}{(\l_1 - \l_2)(\l_1 -                                 \l_3)} +\frac{(d\l_2)^2}{(\l_2-\l_1)(\l_2-\l_3)}  \right. \\
 +  \left. \frac{(d\l_3)^2}{(\l_3-\l_1)(\l_3-\l_2)}\right) +\frac{1}{8\bar{\omega}^2}\left(\frac{\sigma-{\mu}}{\bar{\mu}-\mu} \right)^2 \left( \frac{(d\lb_1)^2}{(\lb_1 - \lb_2)(\lb_1 -                                 \lb_3)}  \right.\\
 +  \left. +\frac{(d\lb_2)^2}{(\lb_2-\lb_1)(\lb_2-\lb_3)}+ \frac{(d\lb_3)^2}{(\lb_3-\lb_1)(\lb_3-\lb_2)} \right)\;.
\label{metric1comb_can}
\end{multline}

The flat coordinates $t$ and $s$ of the metric (\ref{metric1comb_can}) are
\begin{align}
\begin{split}
t =  \frac{\bar{\mu}-\sigma}{\bar{\mu}-\mu} \int_x^{x+2\omega^\prime} \!\!\!\!\! (\wp(\z)+c) \frac{d\z}{2\omega} + \frac{\sigma-{\mu}}{\bar{\mu}-\mu} \int_x^{x+2\omega^\prime} \!\!\!\!\! \overline{(\wp(\z)+c)} \frac{d\bar{\z}}{2\bar{\omega}} \\
s =  \frac{\bar{\mu}-\sigma}{\bar{\mu}-\mu} \int_x^{x+2\omega} \!\!\!\! (\wp(\z)+c) \frac{d\z}{2\omega} + \frac{\sigma-{\mu}}{\bar{\mu}-\mu} \int_x^{x+2\omega} \!\!\!\! \overline{(\wp(\z)+c)} \frac{d\bar{\z}}{2\bar{\omega}} \;.
\label{ts}
\end{split}
\end{align}
We need to perform a linear change of variables in order to have the unit field ${\bf e}$ in the form ${\bf e} = -\d_{t^1}\;.$ After this change of variables, we get the following set of flat coordinates for the metric (\ref{metric1comb_can}): 
\begin{alignat}{2}
\label{flatPhi_comb}
t_1 & := s + \sigma^{-1}t & \qquad t_4 & := s - \sigma^{-1}t \nonumber\\
t_2 & := t^{\scriptscriptstyle{0;1}} = \frac{1}{\omega} \frac{\bar{\mu}-\sigma}{\bar{\mu}-\mu}  
    & \qquad t_5 & := t^{\overline{\scriptscriptstyle{0;1}}} = \frac{1}{\bar{\omega}} \frac{\sigma-\mu}{\bar{\mu}-\mu} \\
t_3 & := r = \frac{1}{2\pi i} \frac{(\bar{\mu}-\sigma)\mu}{\bar{\mu}-\mu}
    &\qquad t_6 & := u = \frac{1}{2\pi i} \frac{1}{\mu - \bar{\mu}} \;. \nonumber
\end{alignat}
In the coordinates (\ref{flatPhi_comb}), the metric has the form:
\beqn
{\bf ds}_{\Phi}^{\bf 2} = \frac{1}{2} (dt_2)^2 + \frac{1}{2} (dt_5)^2 - dt_1 dt_3 + \sigma dt_1 dt_6 - dt_3 dt_4 - \sigma dt_4 dt_6  \;.
\label{sigmametric}
\eeqn
The period of the torus and its complex conjugate can be expressed in terms of the coordinates (\ref{flatPhi_comb}) as follows: $ \mu = t_3/t_6$  and   $ \bar{\mu} = (\sigma - 2 \pi i t_3)/(1 - 2 \pi i t_6) \;,$ respectively.

Then, the prepotential (\ref{prepotential2}) is the following function of $6$ variables:
\begin{align}
 \begin{split}
   & F_{\Phi_s + \sigma\Phi_t}   =  -\frac{1}{64\pi i} \frac{t_2^4}{t_6^2} \gamma \left( \frac{t_3}{t_6} \right) - \frac{\pi i}{16} \frac{t_5^4}{(2 \pi i t_6-1)^2} \gamma \left( \frac{2 \pi i t_3-\sigma}{1-2 \pi i t_6} \right) - \frac{1}{8\pi i} \frac{t_2^2}{t_6}(t_1+t_4)   \\
   & \qquad - \frac{\sigma}{8 \pi i }(t_1^2-t_4^2) + \frac{1}{8\pi i}\frac{t_3}{ t_6}  (t_1+t_4)^2 + \frac{\pi i }{2t_6(t_3-\sigma t_6)} \frac{1}{(2 \pi i t_6 -1)} \times      \\
    & \times \left( \frac{(t_2^2+t_5^2) t_6}{2}  - \frac{t_2^2}{4\pi i}  -  (t_1+t_4) t_3 t_6 + \frac{(t_1+t_4) t_3}{2\pi i} + \sigma (t_1-t_4) t_6^2 - \frac{\sigma (t_1-t_4) t_6}{2\pi i} \right)^2 .
 \label{Fcomb}
 \end{split}
\end{align}
In the limit $\sigma \to 0 \;,$ the metric (\ref{sigmametric}) becomes singular and the function (\ref{Fcomb}) does not satisfy the WDVV system. To obtain from (\ref{Fcomb}) the prepotential $F_{\Phi_s}\;,$ corresponding to the case $\sigma=0\;,$  one has to rewrite $F_{\Phi_s + \sigma\Phi_t}$ in terms of the original variables (\ref{ts}) and then put $\sigma=0\;.$

The function $F_{\Phi_s + \sigma\Phi_t}$ is quasihomogeneous. The quasihomogeneity factors $\{ \nu_i \}$ and $\nu_\iF$ are given by (\ref{quasihomcoeff1}). 

The $G$-function for $\covM_{1;1}^{\Phi_s + \sigma\Phi_t}$ is given by
\beqs
G = - \log \left\{ \eta \left( \frac{t_3}{t_6} \right) \eta \left( \frac{2 \pi i t_3 - \sigma}{1- 2 \pi i t_6} \right) \left( t_2 t_5 \right)^\frac{1}{8} \left( \frac{t_3-\sigma t_6}{t_6(1-2 \pi i t_6)} \right)^{\frac{1}{2}}  \right\} + {\rm const} \;.
\eeqs

A computer check shows that functions $F_{\Phi_s}$ (\ref{Fs}), $F_{\Phi_t}$ (\ref{Ft}), and $F_{\Phi_s + \sigma\Phi_t}$ (\ref{Fcomb}) indeed satisfy the WDVV system.

{\bf Acknowledgments} I am grateful to D. Korotkin, A. Kokotov, M. Bertola and S. Natanzon for many useful discussions and to B. Dubrovin for important comments and pointing out some mistakes in an earlier version of this paper.

\end{document}